\documentstyle[11pt]{article}
\begin{document}
\begin{flushright}
{\sl Preprint} {\bf TUW 96/31 \\ hep-th/9703099} 
\end{flushright}
\begin{center} 
{\Large\bf P-Branes, Poisson-Sigma-Models \\
and Embedding Approach to $(p+1)$-Dimensional Gravity}

\vspace{0.3cm} 

{\large\bf Igor A.\ Bandos$^1$ and Wolfgang Kummer$^2$}

\it{${}^1$  Institute for Theoretical Physics,\\
NSC Kharkov Institute of Physics and Technology, \\
 310108, Kharkov,  Ukraine\\
 e-mail: kfti@rocket.kharkov.ua,\\
 ${}^2$ Institut f\"{u}r Theoretische Physik, \\ Technische
 Universit\"{a}t Wien, \\ Wiedner Hauptstrasse 8-10, A-1040 Wien\\
 e-mail: wkummer@tph.tuwien.ac.at} 
\date{December 1996}
\end{center}

\vspace{1.5cm}

\begin{abstract}
 A generalization of the embedding approach for d-dimensional gravity
 based upon $p$-brane theories is considered.  We show  that the
$D$-dimensional $p$-brane coupled to an antisymmetric tensor field of rank
$(p+1)$ provides the dynamical basis  for the description of $d=(p+1)$
dimensional gravity in the isometric embedding formalism.  ''Physical''
matter appears in such an approach as a manifestation of a $D$-dimensional
antisymmetric tensor (generalized Kalb-Ramond) background.
\\
For the simplest case, the Lorentz harmonic formulation of the
bosonic string in a Kalb-Ramond background and its relation to a
first order Einstein-Cartan approach for $d=2$ dimensional gravity is
analysed in some detail.  A general Poisson-sigma-model structure
emerges.  For the minimal choice of $D=3$ an interesting
``dual'' formulation is found which has the structure of a
Jackiw-Teitelboim theory, coupled minimally to a massive scalar
field.\\
Our approach is intended to serve as a preparation for the study of
 $d$-dimensional supergravity theory, either starting from the generalized action
of free supersymmetric $(d-1)$-branes or $D_{(d-1)}$-branes, or from the
corresponding geometric equations ('rheotropic' conditions).
\end{abstract}

{\bf PACs: 11.15-q, 11.17+y, 04.60.Kz, 04.50.+h, 11.10.Kk.11.30-j.}


\setcounter{page}1
\renewcommand{\thefootnote}{\arabic{footnote}} \setcounter{footnote}0

\newpage

\section{Introduction}

As summarized, e.g.  in ref. \cite{rob}, the embedding approach to gravity has a
long history, beginning at least from the famous book of Eisenhart 
\cite{Ei}.  {\sl ''From time to time, some interesting results would
be derived in this way, but they would also be directly derivable
from the Riemannian metric, interest in the embedding method thus
subsiding again''} \cite{rob}.  Most results obtained by this
approach for General Relativity (GR) can be found in \cite{emb1,Kr}.

 More recently the problems related to the quantization of GR and
some early successes in string theory encouraged Regge and Teitelboim
to study the possibility of a 'string-like' description of gravity
\cite{regge} as an alternative for the 'intrinsic' description in
terms of Riemannian geometry.  This idea generated renewed interest
in the search for a dynamical basis regarding the old embedding
method \cite{emb1,Kr}.  Since then this approach \cite{regge} has
been developed in several papers \cite{pav,maia,tapia,pav1}, 
where e.g.\
in \cite{pav} the model for gravity provided by a free bosonic
p-brane action in curved (but conformally flat) target space-time was
considered.

It should be stressed that at the classical level $p$-brane (and
$D_p$-brane) theories in these approaches have only matter (and gauge
field) degrees of freedom, with gravity involving only auxiliary
non-propagating ones.  But a kinetic term for gravity should appear
in the effective action of the quantum theory after the integration
over the matter fields, yielding propagating gravity as a result of
quantum effects (see, for example, the discussion of this point in
\cite{pav}).  Hence all the embedding models could be regarded as
particular realizations of the concept of gravity induced by quantum
corrections \cite{sakharov,adler}.

 Recent progress to understand some nonperturbative features of
superstring theories \cite{duality,u-dual,str-str,Mth1,Mth2,Mth3,Mth4,Mth5} 
(and refs.\
therein), and especially the concept of '$p$-brane democracy'
\cite{Mth1} adds reasons to search for an adequate description of
Nature in terms of the embedding approach and its supersymmetric
generalization.  Our present paper intends to prepare the ground for
such investigations.

 An important motivation for our study was the observation that a
close relation of the Lorentz-harmonic formulation of bosonic string
theory \cite{bzst,bpstv} exists with the Poisson-sigma-models (PSM)
\cite{strobl,kummer,kummer1}, in which the unification of all types
of matterless $2$-dimensional gravity (including dilaton gravity,
models with dynamical torsion and spherically symmetric 4d gravity)
has been achieved.  For our present purposes we mainly need from that
approach the fact that vielbeine and spin connection are used
systematically as independent variables in a first order formulation. 
We shall demonstrate here that indeed a PSM-like structure appears
naturally in the twistor-like Lorentz harmonic formulation of strings
\cite{bzst,bpstv} after a change of variables corresponding to the
transition to the so-called (Lorentz-) analytical coordinate basis
\cite{sok,niss,wieg,bh,ghs,gds,gst,gikos}.

The general model appears when the string interacting with a
Kalb-Ramond (KR) background is considered.  A key point is that the
interaction with the bosonic string does not put any restriction on
the background at the classical level\footnote{ Equations of motion
for background fields appear only as a condition for the vanishing of
the $\beta$- function of the string, i.e.\ as a condition for
conservation of conformal invariance in quantum string theory
(see \cite{gsw} and refs.\ therein).}.  This is why the interaction
with a KR-background produces an analog of an arbitrary potential
involved in the PSM-action \cite{strobl,kummer,kummer1}, whereas
``matter'' type interactions are induced by the transversal
components of the target space coordinates.

In principle, $D=3$ string theory is enough to describe arbitrary
curved $2$-dimensional space-time.  The deep reason for this is given
by the general theorem about local isometric embedding of analytical
$d=(p+1)$-dimensional manifold into $D$-dimensional flat Minkowski
space-time \cite{Ei,emb,Greene,Kr}. \footnote{See \cite{ak} for a
supersymmetric generalization of the theorem.} It states that such an
embedding is always possible (at least locally) when

 \begin{equation}\label{Dd}
D \geq { {d(d+1)} \over 2}\; .
 \end{equation}
 For $d=2$ ($p=1$) we just have $D=3$.  Hence we can consider
arbitrary $2$-dimensional space-time as bosonic string world-sheet
embedded into flat space time.  The interaction with a KR (second
rank antisymmetric tensor) field makes such an embedding nonminimal. 
The main extrinsic curvatures of the world surface become arbitrary
functions determined by the $D=3$ dimensional KR field strength. 
Thus, such a model describes arbitrary curved $2$-dimensional
space-time and, in this sense, also arbitrary two dimensional
gravity.  In fact, for $D=3$ even in the absence of a KR field, a
complicated PSM structure is found for an action consisting of
``coupled'' versions of the usual first order Einstein-Cartan action
for gravitational theories in $d=2$.  A certain ``duality'' is found
between the zweibeine on the world sheet and of the residual
transverse components of the spin connection.  On the other hand, if
the string is embedded in $D>3$, beside the arbitrary
''matter''contribution provided by the KR field, the minimal coupling
to one ''pre-matter'' scalar field in $D=3$ generalizes to
complicated nonminimal interactions with $D-2$ scalars, $D-2$ one
forms and $(D-2)(D-3)/2$ gauge fields, but maintaining a general PSM
structure, as expected.  The point of departure of our paper,
however, is the generalization of our string model to high
dimensional extended objects ($p$-branes with $p>1$) \cite{bzp,bpstv}
yielding a natural dynamical basis for a description of
$d=(p+1)$-dimensional gravity (in terms of extrinsic geometry) within
the embedding approach \cite{emb1,Kr} \footnote{This result have been
briefly reported in \cite{b1}, where some possible application of the
embedding approach for $F$-theory \cite{Fth1,Fth2,Fth3,Fth4,Fth5} were
considered}.  We find that general $(p+1)$-dimensional gravity is
produced by interaction of the $p$-brane model \cite{bzp,bpstv} with
a generalized KR (GKR) background.

This seems to be a universal property of $p$-brane theory in a GKR
background.  But, to prove this, we need to develop the extrinsic
geometry formalism (i.e.  the so-called geometric approach
\cite{lr,barnes,Zh89,zhelt,bpstv}) for $p$-branes in such a
background.  The Lorentz harmonic formulation \cite{bzst,bzp,bpstv}
is most suitable for this purpose, because it produces the master
equations of the geometric approach (so-called ``rheotropic
conditions'' in the terminology of \cite{bsv}) as equations of motion
(e.o.m.-s) in a straightforward way.

Hence the model can be regarded as a realization of the idea by Regge
and Teitelboim for a 'string-like' description of gravity.  In fact,
some realizations of this idea were proposed previously
\cite{pav,maia,tapia,pav1}.  So in \cite{pav} a model for gravity was
constructed from the free bosonic p-brane action in curved (but
conformally flat) space-time.  \footnote{The extrinsic geometry of
$D=4$ string in a KR field was studied in \cite{lr} (see also
\cite{barnes,Zh89}), as well as in recent work \cite{bakas}.  In
\cite{barnes1} string theory for an anti-de-Sitter background was
investigated and it was demonstrated that the minimality condition
for the embedding is broken in such a case.}

The basic difference of our result with respect to \cite{pav}
consists in the possibility to consider flat target space (i.e.\
without target space gravity), at least at the classical level.  Thus
quantizing that model we bypass the quantum problem of gravity (or
its particular case such as conformal gravity) in the 'large', i.e.\
target space, as well as the embedded, i.e.  world-volume space-time. 
\footnote{From the point of view of supergravity the existence of
both possibilities to describe world-volume gravity, namely by
considering curved target space with nontrivial gravity as well as
the flat one with KR fields only, seems to be very natural.  Indeed,
at high dimensions ($D=10,11$) the only supermultiplet containing the
KR fields is the supergravity multiplet which involves gravity
too.}.  The fact that the KR field may be the origin of arbitrary
gravity in the pure bosonic classical description also seems to be
related to some most recent developments in superstring theory.

The paper is organized as follows: \\
In Section 2 we give a short summary of the embedding method of
General Relativity (GR).  After the metric version we also develop
the one for Cartan variables.  We consider the general case of
$d=(p+1)$ dimensional gravity in a way similar to the one
\cite{bpstv,zero} used in the geometric approach to $p$-branes
(extended objects with $p$ space like dimensions of the world
volume).  Moving frame variables (Lorentz harmonics
\cite{sok,bh,ghs,gds,bzst,bzp,gst,bpstv,bsv}) are introduced here.

The moving frame (Lorentz harmonic) formulation of the $p$-brane
theory \cite{bzst,bzp,bpstv} in the $D$-dimensional generalized
Kalb-Ramond (GKR) background is described in detail in Section 3.  It
is proved that it provides the dynamical ground for $d=p+1$
dimensional gravity in the embedding formalism \cite{emb1,Kr}.

In Section 4 the bosonic string interacting with the KR field is
studied as a model for $d=2$ gravity.  We shortly describe a Lorentz
harmonic formulation of free bosonic strings \cite{bzst,bpstv} and
discuss the PSM-like structure of this formulation appearing in the
world-sheet ('analytical') basis.  The integration with the KR
background shows that a model for a general type of $d=2$ gravity
with matter is obtained in this manner.  As a simplest example we
treat the gravity models inspired by free string theory.

Supersymmetric generalizations of our approach and general directions
of further research are described in the Conclusion.

\section{ The Embedding Method for d-Dimensional Gravity}

\subsection{Metric Approach}

For GR in d dimensional space-time ${\cal M}^d$ is obtained by
specifying the matter, calculating its energy-momentum tensor ${\cal
T}^{mn} (x) \equiv e^m_a e^n_b {\cal T}^{ab} (x)$, and use the latter
as a source for the Einstein field equation

\begin{equation}\label{Einstein}
{\cal R}^{~ac}_{bc} -
1/2 \delta^a_b  {\cal R}^{~dc}_{dc} = 1/2 {\cal T}^a_b (x) \end{equation}

where the curvature two-form is defined by \footnote{Here and below
we use external differential $d$ and external product $\wedge$ of the
forms.  So, $\Omega_r \wedge \Omega_q = (-1)^{rq} \Omega_q \wedge
\Omega_r $, $d(\Omega_r \wedge \Omega_q) = \Omega_r \wedge d\Omega_q
+ (-1)^{q} d\Omega_r \wedge \Omega_q $ for product of any $r$- and
$q$- forms $\Omega_q = dx^{m_q}\wedge ...  \wedge dx^{m_1}
\Omega_{m_1 ...  m_q}(x)$.  As usual $ dd \equiv 0$, in particular
$dd X^{\underline{m}}\equiv 0$.}

\begin{equation}\label{Riem}
{\cal R}^{ab} (d,d) \equiv d\Omega^{ab}
- \Omega^{ac} \wedge \Omega_{c}^{~b}
\end{equation}
$$
\equiv 1/2 dx^m \wedge dx^n {\cal R}^{ab}_{nm}(x)
\equiv 1/2 e^d \wedge e^c {\cal R}^{~ab}_{cd}(x)
$$
 The spin connections $\Omega^{ab}$  are supposed to be the ones,
 constructed from the vielbein fields $(e^a = dx^m  e^a_m) $
$$
e^{m}_{a} \Omega_{{ m bc }} =
$$

\begin{equation}\label{eome} = e^{m}_{a} e^{{n}}_{b} \partial_{{[m}}
e_{{n] c}} - e^{m}_{b} e^{{n}}_{{c}} \partial_{{[m}} e_{{n] a}} +
e^{m}_{c} e^{{n}}_{{a}} \partial_{{[m}} e_{{n] b}} \end{equation}
as a consequence of vanishing torsion

\begin{equation}\label{tor0} T^a \equiv {\cal D} e^a \equiv de^a -
e^b \wedge \Omega^{~a}_b = 0 , 
\end{equation}
with
$$ T^a \equiv 1/2 dx^m \wedge dx^n T_{nm}^{~~a} \equiv 1/2 e^b \wedge
e^c T_{cb}^{~~a} $$ 
the torsion two-form on ${\cal M}^d$.

So the curved space-time ${\cal M}^d$ is described as the solution of
Einstein equation (\ref{Einstein}), (\ref{Riem}), (\ref{eome}),
(\ref{tor0}) for a given matter distribution specified by the
expression for energy-momentum tensor ${\cal T}^{ab}(x)$.

To find exact solutions of the Einstein equations (\ref{Einstein})
for $d=4$ embedding methods were widely used in the past
\cite{emb1,Kr}.  They are based on general theorems \cite{Ei} stating
that any analytical $d$-dimensional curved space-time ${\cal M}^d$
can be considered as a subspace in flat $D$-dimensional
pseudo-Euclidean space-time $R^{1,D-1}$ with $D \leq d(d+1)/2$ at
least locally \footnote{Here we will restrict ourselves by
considering the case of local embedding of analytical spaces only. 
The number of additional dimensions $(D-p)$ being necessary for
global embeddings can be much higher \cite{Kr,emb1}.  Moreover, even
if one replaces the requirement for ${\cal M}^d$ to be analytically
embedded by the requirement of differentiability (${\cal C}^\infty
$), he gets $D \leq d(d+3)/2$ instead of $D \leq d(d+1)/2$
\cite{Greene,maia}.}.  The metric of ${\cal M}^d$ is induced by the
embedding
\begin{equation}\label{metric} ds^2 = dx^m  dx^n g_{mn} (x) = dx^m  dx^n
\partial_n X^{\underline{n}} \partial_m X^{\underline{m}}
\eta_{\underline{n}\underline{m}} ,
\end{equation}
i.e.  it is expressed in terms of derivatives $\partial_m = \partial
/ \partial x^m$ of the coordinate functions defining ${\cal M}^d$
parametrically as the subspace in $R^{1,D-1}$:

\begin{equation}\label{coord}
X^{\underline{m}} = X^{\underline{m}} (x^n) 
\end{equation}
Here $X^{\underline{m}}$ (${\underline{m}} = 0, 1, ..., (D-1)$) denote
Cartesian coordinates of $R^{1,D-1}$, \\ $
\eta_{\underline{n}\underline{m}} \equiv diag (+1,-1,...,-1) $ is the flat
Minkowski metric,  $x^m$ ($m= 0,1,..., (d-1)$) represent local  holonomic
coordinates of  ${\cal M}^d$.

This description of $d$-dimensional space-time is similar to the one
used for the world sheet of strings $(d=2)$ and for the world volume
of p-brane ($d=p+1$) theories.  Further steps towards the description
of the embedding method of GR imply the introduction of an extrinsic
geometry formalism for ${\cal M}^d$.  They are similar to the ones
performed in the geometric approach for strings and $p$-branes
\cite{lr,barnes,bpstv} and will be considered in the same manner
\cite{bpstv}.

\subsection{Extrinsic Geometry in Cartan Variables}

For our present purpose the vielbein formalism turns out to be more
suitable than the metric one.  To understand the embedding conditions
in this case, let us use an appropriate local Lorentz ($SO(1,D-1)$)
transformation to adjust to each point of ${\cal M}^d$ a local moving
(co-)frame of R$^{1,D-1}$

\begin{equation}\label{E}
E^{\underline{a}} =
dX^{\underline{m}} u^{~\underline{a}}_{\underline{m}} , \qquad
\underline{a} =~ 0, 1, ..., D-1 
\end{equation}
by the matrices $ u^{~\underline{a}}_{\underline{m}} = (
u^{~a}_{\underline{m}}, u^{~i}_{\underline{m}}) \in SO(1,D-1)$

\begin{equation}
\label{orthon}
u^{~\underline{a}}_{\underline{m}} \eta^{\underline{m}\underline{n}}
u^{~\underline{b}}_{\underline{m}} =
\eta^{\underline{a}\underline{b}} 
\end{equation}
in such a way that $d=p+1$ vectors $ u^{~a}_{\underline{m}} (x)$ are
parallel to ${\cal M}^d$ and $(D-d)$ vectors
$u^{~i}_{\underline{m}}(x) $ are orthogonal to it:

\begin{equation}\label{Ea}
E^a = dX^{\underline{m}} u^{~a}_{\underline{m}} = e^a , \qquad a =~
0, 1, ..., p \qquad p=d-1
\end{equation}

\begin{equation}\label{Ei}
E^i = dX^{\underline{m}} u^{~i}_{\underline{m}} = 0 , \qquad i =~ 1,
..., (D-d) \qquad
\end{equation}
$e^a = d\xi^m e^{~a}_{m}$ in the r.h.s.\ of (\ref{Ea}) is the
intrinsic vielbein form of ${\cal M}^d$.  Due to (\ref{Ea}) it is
induced by the embedding \footnote{Writing (\ref{Ea}) we identify the
Lorentz group of ${\cal M}^d$ with the $SO(1,d-1)$ subgroup of the
$R^{1,D-1}$ Lorentz group $SO(1,D-1)$ and, in such a way, break
$SO(1,d-1) \otimes SO(1,d-1) \otimes SO(D-d)$ gauge symmetry of the
considered construction up to $SO(1,d-1) \otimes SO(D-d)$.}.

The rectangular blocks
$u^{~a}_{\underline{m}}, ~ u_{\underline{m}}^{i}$ of the
$SO(1,D-1)$-valued moving frame matrix $u^{\underline{a}}_{\underline{m}}
$ are restricted by (\ref{orthon}) which can be decomposed as

\begin{equation}\label{orthon1} u^{~a}_{\underline{m}}
u^{\underline{m}b} = \eta^{ab} \qquad \end{equation}
\begin{equation}\label{orthon2} u^{~a}_{\underline{m}}
u^{\underline{m}j} = 0 \qquad
\end{equation}
\begin{equation}\label{orthon3}
u^{~i}_{\underline{m}}
u^{\underline{m}j} = - \delta^{ij}\; .
\end{equation}
Due to the natural $SO(1,d-1) \otimes SO(D-d)$ gauge invariance of
the present construction, the variables $u^{~a}_{\underline{m}}, ~
u_{\underline{m}}^{i}$ are simply interpreted as homogeneous
coordinates of the coset ${ SO(1,D-1) \over {SO(1,d-1) \otimes
SO(D-d)}}$ .  We call them (vector) Lorentz harmonic variables
\cite{sok} (see also \cite{niss,wieg,bh,ghs,gds,gst}, \cite{bzst,bzp,bpstv}
and refs.\  in \cite{bpstv}) using the term 'harmonic' in the sense of
the work \cite{gikos,sok}.

The form of the induced metric (\ref{metric}) follows from (\ref{Ea}),
(\ref{orthon1}) and the definition $g_{mn} = e^a_m \eta_{ab} e^b_n$.  Eqs.
(\ref{Ea}), (\ref{Ei}) ('rheotropic conditions' \cite{bsv}) are equivalent
to the isometric embedding conditions (\ref{metric}) after the
orthonormality relations (\ref{orthon}) ((\ref{orthon1}) -
(\ref{orthon3})) and the manifest $SO(1,d-1) \otimes SO(D-d)$ gauge
symmetry of Eqs.  (\ref{Ea}), (\ref{Ei}) are taken into account.

Passing from Eqs. (\ref{Ea}), (\ref{Ei}) to their selfconsistency
(integrability) conditions
\begin{equation}\label{dEa} dE^a \equiv
dX^{\underline{m}} \wedge du^{~a}_{\underline{m}} = de^a , \qquad
\end{equation} \begin{equation}\label{dEi} dE^i = dX^{\underline{m}}
\wedge du^{~i}_{\underline{m}} = 0 , \qquad \end{equation} we can exclude
the explicit embedding functions $X^{\underline{m}}(x^n)$ from our further
considerations.

To obtain the differentials of the moving frame variables
$u^{\underline{a}}_{\underline{m}} = ( u^{~a}_{\underline{m}}, ~
u_{\underline{m}}^{i}) $ we shall take into account the
orthonormality conditions (\ref{orthon}) or (\ref{orthon1}) -
(\ref{orthon3}).  Using the unity matrix decomposition

\begin{equation}\label{unity}
\delta^{\underline{n}}_{\underline{m}} = 
 u^{\underline{a}}_{\underline{m}} u^{\underline{n}}_{\underline{a}} = 
 u^{a}_{\underline{m}} u^{\underline{n}}_{a} - u^{i}_{\underline{m}} 
 u^{i\underline{n}} ,
 \end{equation}
being equivalent to (\ref{orthon}) ((\ref{orthon1}) - 
(\ref{orthon3})), the differentials of the moving frame vectors $
u^{\underline{a}}_{\underline{m}} = (u^{a}_{\underline{m}}, 
u^{i}_{\underline{m}})$ become $$ d 
u^{~\underline{a}}_{\underline{m}} = 
u^{~\underline{b}}_{\underline{m}} 
\Omega^{~\underline{a}}_{\underline{b}} (d) \qquad \Leftrightarrow
\qquad $$

 \begin{equation}\label{hdif}
 \cases {
 du^{~a}_{\underline{m}} = u^{~b}_{\underline{m}} \omega^{~a}_b (d) +
u^{~i}_{\underline{m}} f^{ai} (d ) , & \underline{a} = a ; \cr d
u^{i}_{\underline{m}} = - u^{j}_{\underline{m}} A^{ji} +
u_{\underline{m} a} f^{ai} (d) & \underline{a} = i \cr }
\end{equation}
Here

\begin{equation}\label{pC}
 \Omega^{\underline{a}\underline{b}}(d) = -
\Omega^{\underline{b}\underline{a}}(d) \equiv
u^{\underline{a}}_{\underline{m}} d u^{\underline{b}\underline{m}} =
\pmatrix { \omega^{~a}_b (d) & f^{ai} (d ) \cr - f^{bj} (d) & A^{ji}
(d) \cr} 
\end{equation}
is the $so(1,D-1)$ valued Cartan $1$-form.  In (\ref{pC}) it is
decomposed (in a $SO(1,d-1) \otimes SO(D-d)$ gauge covariant way)
into $d \times (D-d)$ covariant forms (forming the basis of the coset
${SO(1,D-1) \over {SO(1,d-1) \otimes SO(D-d)}}$)

\begin{equation}\label{pai}
f^{a~i} \equiv u^{a}_{\underline m} d u^{\underline m~i}\; ,
\end{equation} ${d \times (d-1) \over 2}$ $~~SO(1,d-1)$ connection
1-forms \begin{equation}\label{pab} \omega^{ab} \equiv
u^{a}_{\underline m} d u^{\underline m~b}\; , 
\end{equation}
and (for $D-d ~> 1$), $~~{(D-d) \times (D-d-1) \over 2}$ internal
$SO(D-d)$ connection $1$-forms $A^{ij} (d)$, whose pull-backs onto
${\cal M}^d$, $~~A^{ij} (d) = dx^m A^{ij} _m$, produce world sheet
gauge fields $A^{ij} _m (x)$

\begin{equation}\label{pij}
A^{ij} \equiv u^{i}_{\underline m} d
u^{\underline m~j}\; .
\end{equation}
>From (\ref{pC}) the Cartan forms
$\Omega^{\underline{a}~\underline{b}}$ satisfy the Maurer-Cartan
equation \begin{equation}\label{pMC}
d\Omega^{\underline{a}~\underline{b}} -
\Omega^{\underline{a}}_{~\underline{c}} \wedge
\Omega^{\underline{c}\underline{b}} = 0 \end{equation} which now just
reflect the flatness of the D dimensional embedding space.

The $SO(1,d-1) \otimes SO(D-d)$ gauge covariant splitting of the
connection form $\Omega^{\underline{a}\underline{b}}$ (\ref{pC})
induces the splitting of (\ref{pMC}) into the following set of
equations for the forms $f^{ai},~\omega^{ab}, ~A^{ij}$:

\begin{equation}\label{pPC}
{\cal D} f^{ai} \equiv df^{ai} - f^{bi} \wedge \omega_b^{~a} 
+ f^{aj} \wedge A^{ji} = 0
\end{equation}
\begin{equation}\label{pG}
{\cal R}^{ab} \equiv
d \omega^{ab} - \omega^{ac} \wedge \omega^{~b}_{c} =  f^{ai} \wedge f^{bi}
\end{equation}
\begin{equation}\label{pR}
R^{ij}  \equiv d A^{ij} +
A^{ik} \wedge A^{kj} =  f_a^{i} \wedge f^{aj} 
\end{equation}

Eqs.  (\ref{pPC}), (\ref{pG}), (\ref{pR}) give rise to
Peterson-Codazzi, Gauss and Ricci equations, respectively, of the
subspace embedding theory \cite{Ei,Kr} (see also \cite{bpstv}).

The covariant differential ${\cal D}$ used in Eq.  (\ref{pPC})
include the form $\omega^{ab}$ and $A^{ij}$ as $SO(1,d-1)$ and
$SO(D-p)$ connections.  So its pull-back onto the surface ${\cal
M}^d$ ($~~{\cal D} = dx^m {\cal D}_m = e^a {\cal D}_a$) considered as
a covariant differential on ${\cal M}^d$ implies that spin
connections and gauge fields are being induced by the embedding.

Using this differential we can investigate the integrability conditions
 for Eqs. (\ref{Ea}) and (\ref{Ei}) in manifestly $SO(1,d-1)\otimes
 SO(D-d)$ gauge invariant form

\begin{equation}\label{DEa} {\cal D} E^a
\equiv dX^{\underline{m}} \wedge {\cal D} u^{~a}_{\underline{m}} = {\cal
D} e^a , \qquad \end{equation} 
\begin{equation}\label{DEi} {\cal D}E^i =
dX^{\underline{m}} {\cal D} \wedge u^{~i}_{\underline{m}} = 0  .
\end{equation}
The covariant differentials ${\cal D}$ for the moving frame variables
$ u^{~a}_{\underline{m}}, ~u^{~i}_{\underline{m}} $ from (\ref{hdif})
are expressed in terms of ${ SO(1,D-1) \over {SO(1,d-1) \otimes
SO(D-d)}}$ coset forms $f^{ai}$ alone:

\begin{eqnarray}\label{hDif}
{\cal D} u^{~a}_{\underline{m}} \equiv du^{~a}_{\underline{m}} -
u^{~b}_{\underline{m}} \omega^{~a}_b (d) = u^{~i}_{\underline{m}}
f^{ai} (d ) , \qquad \nonumber\\ {\cal D} u^{~i}_{\underline{m}}
\equiv d u^{i}_{\underline{m}} + u^{j}_{\underline{m}} A^{ji} =
u_{\underline{m} a} f^{ai} (d) .
 \end{eqnarray}
Substituting (\ref{hDif}) into (\ref{DEa}), (\ref{DEi}),  we get
\begin{equation}\label{torp1}
T^a \equiv {\cal D} e^a \equiv de^a - e^b \wedge 
\omega^{~a}_b = E^i \wedge f^{ai} (d)\; .
\end{equation}

\begin{equation}\label{symp1}
0 = {\cal D} E^i \equiv dE^i + E^j \wedge A^{ji} = E_a(d) \wedge f^{ai} (d) ,
\end{equation}
The 'rheotropic conditions' (\ref{Ei}) and (\ref{Ea}) result in the
vanishing of the r.h.s.\ of (\ref{torp1}) and in the possibility to
rewrite the r.h.s.\ of Eq.\ (\ref{symp1}) in terms of the intrinsic
vielbein $e_b$ of ${\cal M}^d$.  Thus, the selfconsistency conditions for
(\ref{Ea}), (\ref{Ei}) imply  the vanishing of torsion (\ref{tor0})
\begin{equation}\label{torp2}
T^a \equiv {\cal D} e^a \equiv de^a - e^b
\wedge \omega^{~a}_b =  0 ,
\end{equation}
and of
\begin{equation}\label{symp2}
e_a(d) \wedge f^{ai} (d) = 0 ,
\end{equation}
respectively.
Eq.\ (\ref{torp2}) determines  the {\it induced} spin connections
$\omega^{ab}$ in terms of the vielbeine. Eq.\ (\ref{symp2}) reflects the
symmetry properties of the second fundamental form matrix \footnote{The
proper definition of the second fundamental form is $K^i \equiv dx^n  dx^m
K_{mn}^i \equiv e^b  e^c K_{ab}^i , ~$ $~K_{mn}^{~~i} \propto \partial_m
\partial_n X^{\underline{m}} \tilde{u}_{\underline{m}}^i $ with
$\partial_n X^{\underline{m}} \tilde{u}_{\underline{m}}^i = 0 $ \cite{Ei}.
} $K^{~~i}_{mn} \equiv e^a_m e^b_n K^{~~i}_{ab} = K^{~~i}_{nm}$ appearing
in the decomposition of the pull-back of the covariant connection form
$f^{ai}$ (see \cite{bpstv}) 

\begin{equation}\label{omek} f^{ai} = e_b
K^{abi}\; .  
\end{equation}
Equations (\ref{pPC})--(\ref{pR}), (\ref{torp2}), (\ref{symp2}) are known
from the Classical Theory of Surfaces \cite{Ei} and describe  an arbitrary
surface embedded into flat $D$-dimensional space-time.

The main extrinsic curvatures of the surface $h^i$ are the traces of the 
second fundamental form matrix $K^{abi}$ 
\begin{equation}\label{mainp}
h^i \equiv K^{abi} \eta_{ab}
= e^m_a f_m^{~ai} \equiv f_a^{~ai} .
\end{equation}
They can be used to define the embedded surface ${\cal M}^d$.
In particular, the world volume of the  free bosonic $p$-brane ($p=d-1$) 
is considered to be a minimal surface, i.e. is defined by $$ h^i = 0 
$$. 
In the embedding method of GR the description of the curved space time ${\cal M}^d$ 
is achieved by the use of the extrinsic geometry equations together with the 
 algebraic  equation
$$
f_c^{~ai}  f_c^{~ci} -
f_b^{~ai} f^{~ci}_c -
1/2 \delta^a_b (
f^{~di}_c f^{~ci}_f -
f^{~di}_d f^{~ci}_c ) \equiv
$$
\begin{equation}\label{Einstein1}
K^{~ai}_c K^{~ci}_b -
K^{~ai}_b K^{~ci}_c -
1/2 \delta^a_b (
K^{~fi}_c K^{~ci}_f -
K^{~fi} _f K^{~ci}_c)
\end{equation}
$$
= {\cal T}^a_b (x) .
$$
This can be derived from the Einstein equation (\ref{Einstein}) when the
Gauss equation (\ref{pG}) is taken into account.

\section{Bosonic P-Branes with Kalb-Ramond Background  and $d=p+1$ Gravity}

The main result of this section is that arbitrary nonvanishing
extrinsic curvature $h^i \neq 0$ can be produced by a $D=
(p+1)(p+2)/2$-dimensional $p$-brane theory interacting with a GKR
background, i.e.  with an antisymmetric tensor field of rank $(p+1)$.

This statement seems to be universal, i.e.  independent of the
special formulation of $p$-brane theory.  But for the proof, the
extrinsic geometry formalism (i.e.  the so-called geometric approach
\cite{lr,barnes,zhelt,bpstv,bsv1,zero}) for $p$-branes in a
GKR-background should be developed.  The Lorentz harmonic formulation
\cite{bzst,bzp,bpstv} is most suitable to do this, because it
produces the master equations of the geometric approach (rheotropic
conditions in the terminology of ref.  \cite{bsv}) as e.o.m.-s (see
below).

\bigskip

We will demonstrate that for a $D$-dimensional $p$-brane interacting
with an antisymmetric tensor field $B_{\underline{m}_1 ... 
\underline{m}_{p+1}} (X^{\underline{m}})$ (GKR field) the main
extrinsic curvatures of the $p$-brane world volume indeed are
nonvanishing

$$
h^i
\equiv K^{~ai}_a \equiv f^{ai} ({\cal D}_a) =
$$
\begin{equation}\label{minipKR}
= -{ 1 \over {(p+1)^2}}
\epsilon _{a_0 ... a_p }
u^{a_0 \underline{m}_0} (x)
...
u^{a_p \underline{m}_p} (x)  \times
\end{equation}
$$
H_{\underline{m}_1
\underline{m}_2 ... \underline{m}_{p+2}} (X(x))\; ,
$$
and expressed in terms of the field strength of the GKR field
$$
H_{\underline{m}_1
\underline{m}_2 ... \underline{m}_{p+2}} (X) \equiv
$$
\begin{equation}\label{KRcstr}
\equiv (\partial_{\underline{m}_1}
B_{\underline{m}_2 \underline{m}_3 ... 
\underline{m}_{p+2}}(X) - \partial_{\underline{m}_2}
B_{\underline{m}_1 \underline{m}_3... \underline{m}_{p+2}}(X) + ...
)\; .
\end{equation}
This means that
\begin{itemize}
\item
any curved space-time
${\cal M}^d$ can be identified locally with  some  world volume of such a type.
 \item
 matter is the manifestation of the GKR  background in such an approach.
 \end{itemize}

\subsection{Action Functional}

Let us consider  a $p$-brane interacting with an antisymmetric tensor gauge 
field of the rank $(p+1)$ (GKR field)

\begin{equation}\label{pKR}
B_{p+1} = dX^{\underline{m}_{p+1}} \wedge ... \wedge dX^{\underline{m}_{1}}
B_{\underline{m}_{1}... \underline{m}_{p+1}} (X) 
\end{equation}
in the  twistor-like Lorentz harmonic  formulation \cite{bzp,bpstv}.

Then the action functional

\begin{equation}\label{acp}
S^{p+1} = S_{0}^{p+1} + S^{p+1}_{int}
\end{equation}
is the sum of the free $p$-brane action \cite{bzst,bzp,bpstv}

\begin{equation}\label{freeac}
 S_0^{(p+1)} = - {1 \over {p!}} \int_{{\cal{M}}^{p+1}} ( E^{a} \wedge
 e^{a_1} \wedge ...  \wedge e^{a_p}
 \end{equation}
 $$
 - { p \over (p+1)} e^{a} \wedge
 e^{a_1} \wedge ... \wedge e^{a_p} ) \epsilon _{a a_1 ... a_p}
 $$
 (which represents the  bosonic limit of the generalized
action for super-$p$-branes \cite{bsv}), and of an interaction term

\begin{equation}\label{intac}
S^{p+1}_{int} = \int_{{\cal{M}}^{p+1}}
B_{p+1}\; .  \end{equation}
In  (\ref{freeac}), (\ref{intac}) the Lagrangian $(p+1)$ forms are
integrated over a world volume ${\cal M}^{p+1}$ of the $p$-brane (which we
will identify with curved $d=p+1$ dimensional space-time) whose local
coordinates are denoted by $\xi^m$ $(a= 0,1,...,p$) (and  will  identified
with the $x^m$ coordinates from the previous section).

For simplicity in (\ref{freeac})  we fix the  ''cosmological constant''
(being proportional to the inverse $p$-brane tension, i.e. to the Regge
slope parameter $\alpha ^\prime $ for the string case) to be equal to one.
$e^a = d\xi^m e^{~a}_{m}$ are world volume vielbeine $(a= 0,1,...,p$),
$E^a $ ($a =~ 0, 1, ..., p$)
are pull-backs of $(p+1)$ $1$-forms (\ref{Ea}) from the target space
vielbein (\ref{E}).  The general target space vielbein $E^{\underline{a}}$
(\ref{E}) is related to a holonomic basis of target space
$dX^{\underline{m}}$ by Lorentz rotations represented by the matrix
(\ref{orthon}).

The variation of the moving frame vectors $u^a ~and ~ u^i$ which does not  
break the orthonormality conditions 
(\ref{orthon}) (or (\ref{orthon1})--(\ref{orthon3})) reduces to
generalized Lorentz ($SO(1,D-1)$) transformations 
$$ \delta
 u^{~\underline{a}}_{\underline{m}} = u^{~\underline{b}}_{\underline{m}} 
 i_{\delta }\Omega^{~\underline{a}}_{\underline{b}} \equiv
 u^{~\underline{b}}_{\underline{m}}  \Omega^{~\underline{a}}_{\underline{b}} 
 (\delta )
 $$
 \begin{equation}\label{hvar}
 \Leftrightarrow \qquad
 \cases {
 \delta u^{~a}_{\underline{m}} = u^{~b}_{\underline{m}} \omega^{~a}_b (\delta ) 
 +  u^{~i}_{\underline{m}} f^{ai} (\delta ) ,
 & {\underline{a}} = {a} ; \cr
 \delta u^{~i}_{\underline{m}} = u_{\underline{m} a} f^{ai} (\delta ) -  
 u^{j}_{\underline{m}} A^{ji} (\delta ),
 & {\underline{a}} = {i}  \cr }
\end{equation}
 where the  parameters of unconstrained variations $i_{\delta}
 \Omega^{~\underline{a}}_{\underline{b}}  =
 \Omega^{~\underline{a}}_{\underline{b}} (\delta ) =  (f^{ai}(\delta ),~
 \omega^{ab}(\delta ), ~A^{ij}(\delta) )$  can be considered as
 contractions of the Cartan forms (\ref{pC}) (or (\ref{pai}) -
 (\ref{pij})) with the variation symbol $\delta $, or equivalently as
 Cartan forms depending on the variation symbol $\delta$ instead of the
 external differential $d$.

Similar expressions for the external differential of the vectors  
$u^{~\underline{a}}_{\underline{m}}$ are obtained from 
(\ref{hdif}) and  are equivalent to the definition of the  connection
forms (\ref{pC}) (or (\ref{pai}), (\ref{pab}), (\ref{pij})).

\subsection{Equations of Motion}

Taking into account the constrained nature of the harmonic variables
(\ref{orthon}) (or (\ref{orthon1}) -- (\ref{orthon3}))
the  variation of the action (\ref{acp}) can be written as
$$
\delta S = \int _{{\cal M} ^{p+1}} [ - { 1 \over {(p-1)!} } (E^a -e^a )
\wedge
e^{a_1} \wedge ...
$$
$$
\wedge e^{a_{p-1}} \wedge
(\delta e^{a_p} - e^b \omega^{~a_p}_b (\delta ) )
\epsilon _{a a_1 ... a_p }
$$
\begin{equation}\label{var}
 - { 1 \over {p!} } f^{ai} (\delta ) \wedge E^i (d)
 e^{a_1} \wedge ...  \wedge e^{a_{p}} \epsilon _{a a_1 ... a_p }
 \end{equation}
 $$
 + {1 \over {p!} } E^i (\delta ) ( f^{ai} (d ) \wedge e^{a_1}
 \wedge ... \wedge e^{a_{p}} \epsilon _{a a_1 ... a_p }
 - p! i_{u^{i}} H_{p+2}  )
 $$
 $$
 - { 1 \over {p!} } {\cal D} E^a (\delta )
e^{a_1 } \wedge  ... \wedge e^{a_{p}}
\epsilon_{a a_1 ... a_p } +
E^a (\delta ) i_{u^{a} } H_{p+2}, 
$$
where
$$
H_{p+2}  \equiv dB_{p+1} =
$$
\begin{equation}\label{pKR1}
 1/(p+2)! dX^{\underline{m}_{p+2}} \wedge ...  \wedge dX^{\underline{m}_1}
 H_{\underline{m}_1 ...  \underline{m}_{p+2}} (X)
 \end{equation} is a GKR
 field strength and
 $$
 i_{c^a} H_{p+2} \equiv
 $$
 \begin{equation}
 { 1 \over
{(p+1)!}} dX^{\underline{m}_{p+2}} \wedge ... \wedge dX^{\underline{m}_2}
c^{\underline{m}_1}
H_{\underline{m}_1 ...  \underline{m}_{p+2}} (X) .
\end{equation} 
The contraction of basic $1$-forms with the 
variation symbol $\delta $ 

\begin{eqnarray}\label{varform} 
E^a (\delta ) \equiv i_{\delta } E^a \equiv \delta X ^{\underline{m}}
u^a_{\underline{m}} \nonumber\\ E^i (\delta ) \equiv i_{\delta } E^i
\equiv \delta X ^{\underline{m}} u^i_{\underline{m}} \nonumber\\
\omega^{ab} (\delta ) \equiv i_{\delta } \omega^{ab} \equiv
u^a_{\underline{m}} \delta u^{b\underline{m}} \nonumber\\ f^{ai}
(\delta ) \equiv i_{\delta } f^{ai} \equiv u^a_{\underline{m}} \delta
u^{i\underline{m}} = \delta u^{a\underline{m}} u^i_{\underline{m}}
\nonumber\\ A^{ij} (\delta ) \equiv i_{\delta } A^{ij} \equiv
u^i_{\underline{m}} \delta u^{j\underline{m}}\; , 
\end{eqnarray}
together with the variation of the world volume vielbein $$
\delta e^a = d\xi ^m \delta e^{~a}_{m}
$$
are taken as a basis in the space of variations.

The absence of the parameters $A^{ij} (\delta )$ in (\ref{var}) reflects
the $ SO(D-p-1)$  gauge invariance of the action (\ref{acp}). The
$SO(1,p)$ gauge invariance manifests itself as the presence of the
parameter $\Omega^{ab} (\delta )$ in the combination $$ \delta e^{a_p} -
e^b \omega^{~a_p}_b (\delta ) $$ only. Hence we can compensate $SO(1,p)$
(pseudo-)\-rotations of harmonic variables  by (pseudo-)\-rotations of the
world volume vielbein $$ \delta_{SO(1,p)} e^{a_p} = e^b \omega^{~a_p}_b
(\delta ) = e^b i_{\delta } \omega^{~a_p}_b \; .  $$ The latter can be
interpreted as world volume Lorentz transformations.  Therefore, the
$SO(1,p)$ subgroup of the target space Lorentz group $SO(1,D-1)$ is
identified as a group of gauge symmetries of the action (\ref{acp}) with
the world volume Lorentz group $SO(1,p)$.

As a consequence,
the world volume spin
connections $w^{ab}$ are singled out which are  induced by the embedding, i.e.
\begin{equation}
w^{ab} = \omega^{ab}\; ,
\end{equation}
where $\omega^{ab}$ is the pull-back of the $SO(1,p)$ connection form
(\ref{pab}).

The variation $E^a(\delta ) \equiv i_{\delta } E^a = \delta
X^{\underline{m}} u^a_{\underline{m}}$ is related to the parameter of
general coordinate (reparameterization) invariance.  For this purpose
the equivalent language of Noether identities is much more
convenient, i.e.\ to analyse the interdependence of the e.o.m.-s as a
result of such gauge symmetries (see below).

Now let us consider the e.o.m.-s which follow from (\ref{var})
in detail starting with the variation of the world volume
vielbeins $\delta e^a$ and the rest of the admissible variations
of moving frame variables characterized by parameter $i_{\delta
} f^{ai}$.  From (\ref{pC}) the Cartan forms \footnote{And
corresponding to the ${{SO(1,D-1)} \over {SO(1,p) \otimes
SO(D-p-1)}}$ (''boost'') transformations} lead to the
nondynamical equations (\ref{Ea}), (\ref{Ei}) which are the
master equations of the geometric approach
\cite{lr,barnes,bpstv} as well as for the embedding method of GR
\cite{emb1,Kr} (cf.\ section 2):  From $\delta e^a$ follows $$
(E^a -e^a ) \wedge e^{a_1} \wedge ...  \wedge e^{a_{p}} \epsilon
_{a a_1 ...  a_p } = 0 $$ or

\begin{equation}\label{rhEa} E^a \equiv dX^{\underline{m}}
u^a_{\underline{m}} = e^a , 
\end{equation} 
whereas from $i_{\delta }
f^{ai} = f^{ai}(\delta)$ we obtain $$ E^i \wedge e^{a_1} \wedge ... 
\wedge e^{a_{p-1}} \epsilon _{a a_1 ...  a_p } = 0 $$ or

\begin{equation}\label{rhEi}
E^i \equiv dX^{\underline{m}} u^i_{\underline{m}} = 0 \; . 
\end{equation}
Eqs.\ (\ref{rhEa}) and (\ref{rhEi}) precisely lead to the relations used
in section 2:  The pull-backs of the $(p+1)$ basic one forms $E^a$ of the
target space become tangent to the world volume and coincide with world
volume vielbeine $e^a$ (which in turn are  introduced  by embedding on the
shell of the rheotropic conditions), and the pull-backs of the remaining
$(D-p-1)$ basic $1$-forms $E^i$ vanish. Thus  the $(D-p-1)$ vectors being
dual to these $1$- forms become orthogonal to the world volume.

As demonstrated in section 2, the selfconsistency conditions for the
rheotropic relations (\ref{rhEa}), (\ref{rhEi}) are identical to
eqs.\ (\ref{torp2}), (\ref{symp2}), respectively.

The covariant differential ${\cal D}$
appearing in eqs.\ (\ref{torp2}) and (\ref{symp2})  is the pull-back of
the one defined in Eq.\  (\ref{hdif}) (i.e.\  with $SO(1,p)$ and
$SO(D-p-1)$ connections induced by embedding $w^{ab} = \omega^{ab}$,
$B^{ij} = A^{ij}$).

The variation  $ E^a (\delta) \equiv i_{\delta }E^{a} = \delta
X^{\underline{m}} u^a_{\underline{m}}$ does not lead to independent
e.o.m.-s.  In accordance with the second Noether theorem this means that
$E^a (\delta )$ is related to $(p+1)$ parameters of the gauge symmetry of
the $p$-brane action, namely reparametrization symmetry (or general
coordinate invariance for the world volume).

This can be seen from the e.o.m.\ for $E^a(\delta)$ \footnote{The
input from the surface term appearing due to the integration by
parts is neglected here for simplicity.  Hence we treat the case of
closed $p$-branes only.}

\begin{equation}\label{varEa}
(-1)^{p+1} {1 \over {(p-1)!}} {\cal D} e^a \wedge e^{a_1} \wedge ... 
\wedge e^{a_{p-1}} \epsilon _{a a_1 ...  a_p } = i_{u^a} H_{p+2} \; .
\end{equation}
The l.h.s.\  of Eq.\
(\ref{varEa}) vanishes due to (\ref{symp2})  and the r.h.s.\ $$
i_{u^{\underline{m}}_a} H_{p+2}
= { 1 \over  {(p+1)!}} dX^{\underline{m}_{p+1}} ...  dX^{\underline{m}_1}
 u^{\underline{m}}_a
 H_{{\underline{m}} {\underline{m}}_1 ...
  {\underline{m}}_{p+1} } (X(\xi)), $$
is proportional to $$ \propto
 u^{\underline{m}_{p+2}}_{a_{p+2}}
 ...
 u^{\underline{m}_{2}}_{a_{2}}
 u^{\underline{m}}_a
  H_{{\underline{m}} {\underline{m}}_2 ... {\underline{m}}_{p+2} } ,  $$ 
 on the surface of the rheotropic conditions (\ref{rhEa}), (\ref{rhEi}) 
 (where $dX^{\underline{m}} = e^a u_a^{\underline{m}} $).  The latter 
 expression vanishes also identically as an antisymmetric  tensor of  rank
 $(p+2)$ with $(p+1)$ valued vector indices.

We emphasize the interesting  fact that all the independent equations 
considered above ((\ref{rhEa}), (\ref{rhEi})) remain the same for the case 
of a free $p$-brane theory.

In the following we drop  the rheotropic conditions (\ref{rhEa}), 
(\ref{rhEi}) and consider only their integrability conditions 
(\ref{torp2}), (\ref{symp2}) together with  (\ref{pMC}) (equivalent to 
((\ref{pPC}) -- (\ref{pR})) which can be solved by expressions (\ref{pC}) 
(or (\ref{pai}) -- (\ref{pij}))  for connection forms  in terms of moving 
frame variables.

The embedding of an
arbitrary $(p+1)$-dimensional subspace into flat space-time of dimension 
$D \geq (p+1)(p+2)/2$ is encoded in Eqs.\ 
(\ref{torp2}), (\ref{symp2}), (\ref{pPC})--(\ref{pR}).

To further specify the subspace under  consideration it is necessary to  
define its main extrinsic curvatures
\begin{equation}
h^i \equiv \eta_{ab}
 K^{bai}\; ,
 \end{equation}
i.e.\ the traces of the second fundamental
form matrix.  This is just implied  by the last e.o.m.\ following from 
$i_{\delta }E^i = E^i(\delta)$ in (\ref{var}),
$$
f^{ai} \wedge
e^{a_1} \wedge  ... \wedge  e^{a_{p}}
\epsilon _{a a_1 ... a_p } =
$$
\begin{equation}\label{varEi}
= -{1 \over {p!}}
h^{i}
e^{a_0} \wedge  ... e^{a_{p}} \wedge
\epsilon _{a_0 ... a_p } =
- p! i_{u^i} H_{p+2}
\end{equation}
Thus, the main extrinsic curvatures
$$
h^i
\equiv K^{~ai}_a \equiv f^{ai} ({\cal D}_a) =
$$
\begin{equation}\label{varEi1}
=- { 1 \over {(p+2)^2}}
\epsilon _{a_0 ... a_p }
u^{a_0 \underline{m}_0} (x)
...
u^{a_p \underline{m}_p} (x) \times
\end{equation}
$$
\times H_{\underline{m}_1
\underline{m}_2 ... \underline{m}_{p+2}} (X(x))
$$
 of the embedded world volume are defined in terms  of the $D$-dimensional
 GKR field strength  which may be regarded as an arbitrary function of the
 coordinate functions $X^{\underline{m}}(\xi) $.  This means that our 
model --- at least locally ---  provides the basis for a description of arbitrary 
curved space-time of dimension $d$ satisfying $D \geq d(d+1)/2 $  as a 
subspace in $D$-dimensional flat space-time.

 Of course, to solve these equations in the general case some kind of 
 'selfconsistency field technique' need be developed.

 From the point of view of recent developments in string theory, the most 
 interesting ones are possibly related to systems with a solitonic 
 solution similar to the generalized magnetic monopoles \cite{nepomechie} 
 as a source for generating the main curvature. Further investigations 
along this line seem to be promising.

\section{Bosonic String in KR Background and Two-Dimensional Gravity}

 We now treat  the simplest nontrivial example of the models proposed 
 above. Of course most results follow from the general considerations of 
 Section 3. But we take the opportunity to introduce a light-like basis, 
 most useful in $d=2$, and to also fix some notations for later purposes. 
 In an analytical basis of the target space (\ref{pKR})  we find a close 
 relation to PSM-type models \cite{strobl,kummer,kummer1} after 
 incorporation of the Maurer Cartan equations into the action by Lagrange 
multipliers.

For the free string a suitable change of variables leads to the
so-called Jackiw-Teitelboim model with an intriguing
'picture-duality' property emerging for such strings in $D=3$.

\subsection{Twistor-like Action}

In light-like notation the action
describing the bosonic string interacting with  a KR background reads
$$
S = - 1/2
\int_{{\cal M}^2} ~~ ( E^{+}(d) e^{-} (d) - E^{-}(d) e^{+}(d)
$$
\begin{equation}\label{acstr}
- e^{+}(d) e^{-}(d)) + \int_{{\cal M}^2} ~~~ B_{2} (d,d)\; ,
\end{equation}
 where the second term involves the KR field $B_{\underline{n}
 \underline{m}}(X(\xi ))$
 $$
 B_2 \equiv B_2
 (d,d) \equiv 1/2 dX^{\underline{m}} dX^{\underline{n}} B_{\underline{n}
  \underline{m}}(X)
 $$
 \begin{equation}\label{2form}
  = d\xi^m d\xi^n \partial_m X^{\underline{m}}
\partial_n X^{\underline{n}} B_{\underline{n} \underline{m}} (X(\xi))\; .
\end{equation}
${\cal M}^2$ is a bosonic world sheet with local coordinates $\{
\xi^m \} = \{ \tau , \sigma \}$, $e^{\pm} = d\xi^{m} e_m^{\pm} \equiv
e^0 \pm e^1 $ are the covariant light cone components of the zweibein
$1$-form of the string world sheet.  The signs in the superscript
($+$ and $-$) denote the weight with respect to the action of the
world sheet Lorentz group $SO(1,1)$.

\begin{equation}\label{E+def}
E^{+} = d X^{\underline{m}} u_{\underline{m}}^{+} , \qquad 
\end{equation}
\begin{equation}\label{E-def}
E^{-} = d X^{\underline{m}}
u_{\underline{m}}^{-} , \qquad
\end{equation}
are the pull-backs (onto the world sheet) of two basic 1-forms of
target space-time written in the light-like notation  
$$ E^a = (1/2\,
(E^{+}+E^{-}), 1/2\, (E^{+}+E^{-})) , \qquad a= 0,1\; .  
$$
Together with the other $(D-2)$ forms $E^i$ (\ref{Ei}) 
$$
E^{i} = d X^{\underline{m}} 
u_{\underline{m}}^{i} , \qquad
$$
they form the basis (\ref{E}) 
 of the space cotangent to target space-time.  This basis differs 
from the holonomic basis $dX^{\underline m}$ by $SO(1,D-1)$ Lorentz 
transformations represented by the components of the orthogonal 
moving frame matrix (\ref{orthon}) which reads here 

\begin{equation}\label{split}
u_{\underline{m}} ^{\underline{a}}
 \equiv (1/2(u_{\underline{m}}^{+}+
 u_{\underline{m}}^{-}), u_{\underline{m}}^i, 1/2 (u_{\underline{m}}^{+}  - u_{\underline{m}}^{-}))\; .
 \end{equation}
 The vectors $u^{\pm}, u^i$ can be identified with  the ${{SO(1,D-1)} \over {SO(1,1) \otimes SO(D-2)}}$  (vector) Lorentz harmonics
 \cite{sok,niss,bpstv,zero} (cf.\  section 2).
In the present case of $d = 2$
 the orthonormality conditions (\ref{orthon1}) - (\ref{orthon3}) become
\begin{equation}\label{orthog}
u_{\underline{m}}^{+} u^{\underline{m}+} = 0 , \qquad  u_{\underline{m}}^{-} u^{\underline{m}-} = 0 , \qquad
u_{\underline{m}}^{i}
 u^{\underline{m}\pm} = 0 , \qquad
\end{equation}
\begin{equation}\label{norm}
 u_{\underline{m}}^{-}
 u^{\underline{m}+} = 2 ,
\qquad
 u_{\underline{m}}^{i}
 u^{\underline{m}~j} = - \delta^{ij} ,
 \end{equation}
the coset forms $f^{ai}$ (\ref{pai}) are decomposed into two covariant
$SO(D-2)$ vector forms
\begin{equation}\label{+i} f^{+i} \equiv
u^{+}_{\underline m} d u^{\underline{m}i} 
\end{equation}

\begin{equation}\label{-i}
 f^{-i} \equiv u^{-}_{\underline m} d u^{\underline{m}i} , 
\end{equation}
the $SO(1,1)$ connection $\omega^{ab}$ (\ref{pab}) having the  simple
representation $\omega^{ab} \propto \omega \epsilon^{ab}$

\begin{equation}\label{0}
\omega \equiv {1 \over 2} u^{-}_{\underline m} d
 u^{\underline  m~+}\; ,
\end{equation}
and $SO(D-2)$   connections
('gauge fields') are defined by (\ref{pij})

\begin{equation}\label{ij}
 A^{ij} \equiv u^{i}_{\underline m} d u^{\underline m~j}\; .
\end{equation}
The Maurer-Cartan equations (\ref{pMC})
split naturally in terms of
$(\ref{+i})--(\ref{ij})\;$:

\begin{equation}\label{PC+} {\cal D} f^{+i}
\equiv d f^{+i} -  f^{+i} \wedge \omega + f^{+j} \wedge A^{ji} = 0  
\end{equation}
\begin{equation}\label{PC-}
 {\cal D} f^{-i} \equiv d  f^{-i} +  f^{-i} \wedge \omega +  f^{-j} \wedge A^{ji} = 0
\end{equation}
\begin{equation}\label{G}
 {\cal R} \equiv d \omega = {1\over 2} f^{-i} \wedge f^{+i}  
\end{equation}
\begin{equation}\label{R}
f^{ij}  \equiv d A^{ij} + A^{ik} \wedge A^{kj} = - f^{-[i} \wedge f^{+j]} 
\end{equation}
Thus the variations of  moving frame vectors $
\delta u^{~\underline{a}}_{\underline{m}} = u^{~\underline{b}}_{\underline{m}} i_{\delta } \Omega^{~\underline{a}}_{\underline{b}}  = u^{~\underline{b}}_{\underline{m}}  \Omega^{~\underline{a}}_{\underline{b}} (\delta) $, being
necessary for the derivation of the e.o.m.-s of the action
(\ref{acstr}), become
\begin{equation}\label{varh}
\delta  u^{\pm}_{\underline{m}} = \pm 1/2 u^{\pm}_{\underline{m}} \omega
(\delta) + u^{i}_{\underline{m}} f^{\pm ~i} (\delta) ,
\end{equation}
The nondynamical equations of motion ('rheotropic
conditions'\cite{bsv}), which appear as a result of variations with
respect to the moving frame variables ($i_{\delta } f^{ai} \equiv
f^{ai}(\delta )$) and the vielbeine $e^{\pm}$

\begin{equation}\label{ei}
 E^i= 0
\end{equation}
\begin{equation}\label{e+}
E^{+} = e^{+}
\end{equation}
\begin{equation}\label{e-}
E^{-} = e^{-}\; ,
\end{equation}
could  be collected in one equation
 \begin{equation}\label{dX}
 dX^{\underline{m}} = 1/2 e^{\mp}
u^{\pm \underline{m}}
\end{equation}
which  can be solved, in
 particular, with respect to $u^{\pm \underline{m}}$. Substituting this
 expression into Eqs.  (\ref{+i}), (\ref{-i}) we find that the coset forms
 $f^{\pm i}$ are  related to the second fundamental form  of the embedded
 surface $$ K^{~~i}_{mn} \propto \partial_m \partial_n X^{\underline{m}}
u^{~i}_{\underline{m}} \qquad ~~~ ( \partial_n X^{\underline{m}}
u^{~i}_{\underline{m}} = 0 ) $$ (whose traces define the main curvatures
\cite{Ei}) by \begin{equation}\label{second} f^{\pm i} = d\xi^m
\Omega_m^{\pm i} \qquad \Omega_m^{\pm i} \propto  K^{~~i}_{mn} e^{n \pm}
\; .  \end{equation}

The evident symmetry of the second fundamental form
$K^{~~i}_{mn} = K^{~~i}_{nm}$ with respect to permutations of
the world sheet vector indices implies that the $f^{\pm i}$
satisfy

\begin{equation}\label{symm} f^{-i}\wedge e^{+} + f^{+i}\wedge e^{-}
= 0 \qquad \Rightarrow \qquad f^{-i}_{-} = f^{+i}_{+} \equiv 1/2 h^i
\end{equation} which indeed also follows from selfconsistency
conditions of the rheotropic relations (\ref{ei}).  On the other
hand, (\ref{e+}) and (\ref{e-}) yield vanishing world sheet torsion
\begin{equation}\label{tor} T^{\pm} \equiv {\cal D} e^{\pm} = d
e^{\pm} \mp e^{\pm} \omega = 0 \; .  
\end{equation}

The only proper dynamical equation appears as a result of the
variation with respect to the $X$ variable.  Its independent
part

$$
f^{+i} \wedge e^{-} - f^{-i} \wedge e^{+} \equiv e^{+} \wedge e^{-}~h^i~~
= 2 i_{u^{i}} H_3
$$
\begin{equation}\label{minKR}
\equiv dX^{\underline{m}}(\xi) \wedge dX^{\underline{n}}(\xi)
u^{i\underline{l}} H_{\underline{l} ~\underline{n} ~\underline{m}}
(X(\xi)) 
\end{equation}
can be extracted by contraction with $u^{i}_{\underline{m}}$ and
expresses the main curvature of the world sheet
\begin{equation}\label{minKR1} h^i \equiv f^{-i}_{+} + f^{+i}_{-} = 2
f^{+i}_{+} = { 1 \over {2}} u^{\underline{m}i}
u^{\underline{n}-}u^{\underline{k}+}
H_{\underline{k}~\underline{n}~\underline{m}}(X(\xi )) 
\end{equation}
through the pull-back on the KR field strength

\begin{equation}\label{KR}
 H_3 \equiv dB_2 = 1/3! dX^{\underline{m}} \wedge dX^{\underline{n}}
\wedge dX^{\underline{k}} H_{\underline{k}\underline{n}
\underline{m}}\; .  
\end{equation}
Thus, the main curvature of the world volume can be assumed to be an
arbitrary function of the world volume coordinates by appropriately
choosing the KR field.  In accordance with the general theorem
\cite{emb1,Ei} mentioned in the previous Section, this means that
this model even describes arbitrary two-dimensional surfaces already
for the case of $D=3$ target space dimensions.

\subsection{Bosonic String and Poisson-Sigma-Models}

During recent years it  became evident that {\em all} 2d covariant models
of gravity, including generalized dilaton models and models with gauge
fields etc. are special cases of so-called Poisson-Sigma-Models (PSM)
\cite{strobl,kummer}
\begin{equation}\label{PsMg}
{\cal S}_{PSM} =
\int_{{\cal M}^2} ( Y^{\cal C} d{\cal A}_{\cal C} + {\cal P}^{{\cal
B}{\cal C}}(Y) {\cal A}_{\cal B} \wedge {\cal A}_{\cal C}) 
\end{equation}
where ${\cal P}^{{\cal B}{\cal C}}(Y)$ is a Poisson-structure on
${\cal M}^2$, the (zero-forms) $Y$ represent PSM target space
coordinates, ${\cal A}_{\cal B}$ are one-forms.  Generally valid
properties like the existence of absolute conservation laws
\cite{strobl,kummer2}, special behaviors of quantized systems of this
type \cite{strobl,37a}, etc.\ are well-known by now.
Here we shall argue that that the considered bosonic string action
acquires the structure of an action functional for $d=2$ gravity in
the PSM approach.  This structure becomes evident after a change of
variables is carried out and an integration by parts is performed. 
We recall that the PSM-action for $d=2$ gravity has the form of first
order Einstein-Cartan theory \cite{kummer}
  $$  S_{PSM} =  $$

\begin{equation}\label{Psm}
 = -  1/2 \int \Big( \tilde{X}^{+}  \tilde{{\cal D}} e^{-} -
 \tilde{X}^{-} \tilde{{\cal  D}} e^{+} + \tilde{X}^{\perp} d
 \tilde{\omega}
 \end{equation}
 $$
 - e^{+} e^{-}  V(\tilde{X}^{\perp},\tilde{X}^{+}
 \tilde{X}^{-}) \Big)
 $$
  where $X^{\pm}, \tilde{X}^{\perp}$
 are Lagrange multipliers, $\tilde{\omega}$ is a world sheet spin
 connection (regarded as independent 1-form variable) and,  defining a
 covariant differential $\tilde{\cal D}$, $$ \tilde{\cal D} e^{\pm} = d
e^{\pm} \mp e^{\pm} \wedge \tilde{\omega}\; .  $$ $V( \tilde{X}^{\perp},
X^+ X^-)$ is an arbitrary function.  By the e.o.m.\ for $\delta X^\pm$ the
derivative of $V$ defines world-sheet torsion

\begin{equation}\label{torpsm}
\tilde{T}^{\pm} \equiv
\tilde{\cal D} e^{\pm} = d e^{\pm} \mp e^{\pm} \tilde{\omega} = e^{+}
e^{-}{{\partial V} \over {\partial \tilde{X}^{\pm}}}\; , 
\end{equation}
and from $\delta \tilde{X}^\perp$ Riemannian curvature

\begin{equation}\label{Psmcur}
R \equiv {}^{*} d \tilde{\omega} =
{\partial V(\tilde{X}^{\perp },\tilde{X}^{+} \tilde{X}^{-})
\over {\partial \tilde{X}^{\perp }}}
\end{equation}
By elimination of $X^\perp$ and $X^\pm$ evidently any $2d$-theory
with arbitrary powers in curvature and torsion can be constructed. 
By a conformal transformation $e^{\pm} = e^{\phi (\tilde{X}^{\perp
})} \tilde{e}^{\pm}$ also the generalized dilaton theories can be
obtained, comprising, e.g.\ spherically reduced gravity \cite{37b}
and the string-inspired dilaton black-hole \cite{37c}.

Recently it has been proved \cite{kummer1} that $d=2$ gravity with
torsion locally is equivalent to torsionless generalized dilaton gravity. 
Hence, we restrict our considerations to cases with a
potential $V$ being independent of $ X^\pm$, i.e.\ $ V = V(
\tilde{X}^{\perp})\; .$ \\ In order to exhibit a similar structure in
the bosonic string formulation (\ref{acstr}) a change of variables in
the free part of the string action

\begin{equation}\label{anal} (X^{\underline{m}}; u^{\pm}_{\underline{m}},
u^{i}_{\underline{m}}) \qquad \rightarrow \qquad  (X^{\pm}, X^i;
u^{\pm}_{\underline{m}}, u^{i}_{\underline{m}})  
\end{equation} 
$$ X^{\pm} \equiv X^{\underline{m}} u^{\pm}_{\underline{m}} , \qquad
X^i \equiv X^{\underline{m}} u^{i}_{\underline{m}} $$ should be made. 
It corresponds to a transition to an analytical basis
\cite{gikos,sok,bh} of the target Minkowski space-time.  The
pull-backs of the target space vielbein forms $E^{\pm}$ are expressed
in terms of these variables are expressed as

\begin{equation}\label{Ean}
 E^{+} = {\cal D} X^{+} - X^i f^{+i}, \qquad  E^{-} = - {\cal D} X^{-}
 + X^i f^{-i} \; .
 \end{equation}
In (\ref{Ean}) ${\cal D}$ is the world sheet covariant derivatives
involving the pull back of the Cartan form $\omega$ (\ref{0}) as the spin
connection (this is possible due to the identification of the $SO(1,1)$
subgroup of the target space structure group $SO(1,D-1)$ with the world
sheet Lorentz group, which corresponds to  the property of the Lorentz
harmonic formulation \cite{bzst,bzp,bsv}) :

\begin{equation}\label{der}
{\cal D} X^{+} \equiv dX^{+} -  X^{+} \omega , \qquad {\cal D} X^{-} 
\equiv dX^{-} + X^{-} \omega , 
\end{equation}
For fields with $SO(D-2)$ indices the pull-backs of (\ref{ij}) are used as
''internal'' $SO(D-2)$ connections in ${\cal D}$
\begin{equation}\label{ider} {\cal D} X^{i} \equiv dX^{i} + X^{j} A^{ij}\; . 
\end{equation}
Substituting expressions (\ref{Ean}), (\ref{der}) into Eq. (\ref{acstr})
and integrating by parts (neglecting the surface terms for simplicity
\footnote{So, we consider closed string theory here.}) we obtain the
action functional for the bosonic string in the  form $$ S = - 1/2 \int
\Big( X^{+} {\cal D} e^{-} - X^{-} {\cal D} e^{+} + $$

\begin{equation}\label{acanf}
+ X^i (f^{-i} \wedge e^{+}- f^{+i} \wedge e^{-}) - e^{+} \wedge e^{-}
\Big) + \int B_2 \qquad .  \end{equation} Extracting the world sheet
 volume two-form  $e^{+} \wedge e^{-}$ in the last two terms we arrive at
$$
 S = - 1/2 \int \Big( X^{+} {\cal D} e^{-} - X^{-} {\cal D} e^{+} +
\qquad  $$

\begin{equation}\label{acan}
 +   e^{-} e^{+} [1+  X^i ( (f^{-i}_{-} + f^{+i}_{+})] \Big) + \int B_2
\end{equation}
which is similar to but not  identical to the PSM-action (\ref{Psm}).

The first (free) part of this action involves the target space
coordinate fields $X^{\pm}, X^i$ (the latter with $SO(1,D-1)$ Lorentz
indices) and the pull-backs of one-forms

\begin{equation}\label{fpmi1}
f^{\pm i} = e^{+} f^{~\pm i}_{+} + e^{-} f^{~\pm i}_{-} =d \xi^{m}
f^{~\pm i}_{m} 
\end{equation}

\begin{equation}
\omega = e^{+}  \omega_{+} + e^{-} \omega_{-} =
d \xi^{m} \omega_{m}
\end{equation}
only.  In this approach the latter variables are still constrained to
satisfy (\ref{pMC}) (or equivalently (\ref{PC+}) - (\ref{R})), i.e.\
we do not need any reference to the explicit expressions
(\ref{pC}) (or (\ref{+i}) - (\ref{ij})) for them in terms of harmonic
variables.  Varying the action with respect to $f$ of (\ref{fpmi1}) and
$\omega$  this must be taken into account.  $\delta
\Omega^{\underline{a}\underline{b}}(d)$ can be determined from
(\ref{pMC}) contracted with the variation symbol,

\begin{equation}
 i_{\delta} (d \Omega^{\underline{a}\underline{b}}  -
\Omega^{\underline{a}}_{~\underline{c}} \wedge
\Omega^{\underline{c}\underline{b}}) = 0 ,
\end{equation}
i.e.
\begin{equation}\label{MCvar}
\delta \Omega^{\underline{a}\underline{b}} (d) = d
\Omega^{\underline{a}}_{~\underline{b}} (\delta ) +
\Omega^{\underline{a}\underline{c'}} (\delta)
\Omega_{\underline{c'}}^{~\underline{b}} (d) -
\Omega^{\underline{a}}_{~\underline{c'}} (d)
\Omega^{\underline{c'}\underline{b}} (\delta) \; , 
\end{equation} 
and thus

\begin{eqnarray}\label{fomAvar}
\delta f^{\pm i}& =& {\cal D} f^{\pm i} (\delta ) - f^{\pm i} \omega
(\delta ) + f^{\pm j} A^{ji}(\delta ),\nonumber \\ \delta \omega &=&
d\omega (\delta ) - 1/2 f^{- i}f^{+ i}(\delta ) + 1/2 f^{- i}(\delta
)f^{+ i}, \nonumber \\ d A^{ij} &=& {\cal D} A^{ij}(\delta ) - f^{-
[i}f^{+ j]}(\delta ) + f^{-[i}(\delta )f^{+j]} \quad .
\end{eqnarray}
The independent variations are produced by the contractions of the 
corresponding components of the spin connection one-forms with the 
variation symbol $\delta$

\begin{equation}\label{basevar}
 i_{\delta} f^{\pm i} = f^{\pm i} (\delta ), \qquad i_{\delta} 
 \omega = \omega (\delta ), \qquad
{\rm i}_{\delta} A^{ij} =  A^{ij}(\delta ) ,
\end{equation}
which parameterise the Lorentz group algebra $so(1,D-1)$.

To vary the second term of the action (involving the KR field which
is assumed to be dependent on the coordinates $X^{\underline{m}}(\xi
)$ only) in the analytical basis we use the expression for the
variation of the differential form and completeness of the set of
moving frame variables

$$
\int \delta B_2 = \int (~i_{\delta} ~(dB_2) + ~d (i_{\delta}~B_2)) =
$$
\begin{equation}\label{varB2}
= \int ~i_{\delta} ~(dB_2) =
 \int 1/2 dX^{\underline{m}} \wedge dX^{\underline{n}} \wedge
\delta X^{\underline{k}}
H_{\underline{k}\underline{n}\underline{m}} =
\end{equation}
$$
=
\int 1/2 E^{\underline{a}} \wedge  E^{\underline{b}} \wedge
E^{\underline{c}} (\delta ) H_{\underline{c}\underline{b}\underline{a}}
\qquad
$$
 The forms
$E^{\pm}, E^{i}$ and their contractions with the variation symbol $$
i_{\delta} E^{\underline{a}}  \equiv E^{\underline{a}} (\delta ) \equiv
(E^a (\delta), E^{i}(\delta ))
= \delta X^{\underline{m}} u^{\underline{a}}E_{\underline{m}} $$
are now expressed
in terms of the analytical basis coordinates by (\ref{Ean})
and
\begin{equation}\label{Eian} E^{i} = {\cal D} X^{i} - 1/2 X^{+} f^{+i}  -
 1/2  X^i f^{-i}\; . 
\end{equation}
The action (\ref{acan})  expressed in that basis contains  $X^{\pm}$ which
play the role of  Lagrange multipliers for the condition of vanishing
torsion for the induced connection $\omega$ (\ref{tor})
\begin{equation}\label{toran}
 T^{\pm} \equiv {\cal D} e^{\pm} = d  e^{\pm} \mp  e^{\pm} \wedge  \omega = 0 ,
\end{equation}
just as in Einstein-Cartan action (\ref{Psm})
 \cite{strobl,kummer,kummer1} for  torsionless gravity.

 The field $X^i$ is a Lagrange multiplier for the condition defining the main curvatures
 of the world sheet.
In the case of the free string ($B_2 = 0$) they simply vanish
$$
h^i \equiv f^{-i} ~~e^{+} - f^{+i} ~~e^{-} = 0 \qquad \Rightarrow
\qquad
$$
\begin{equation}\label{min}
h^i \equiv f^{-i}_{-} + f^{+i}_{+} = 0  
\end{equation}
and this means that the world sheet is embedded into the flat
$D$-dimensional Minkowski space-time as a minimal surface.

In the corresponding case of a
nonvanishing KR background one obtains in the analytical basis by
(\ref{varB2}),  (\ref{minKR}) or (\ref{minKR1})
\begin{equation}\label{minKR2}
 h^i \equiv f^{-i}_{+} + f^{+i}_{-}  =
 2 f^{+i}_{+} = { 1 \over {2}}
H^{+-i}\; ,
\end{equation}
where the function $H^{+-i}$
$$
H^{+-i} = u^{\underline{m}i} u^{\underline{n}-}u^{\underline{k}+}
H_{\underline{k}~\underline{n}~\underline{m}}(X^{\underline{l}}(\xi)) 
$$
depends not only on the coordinates
 $X^{\pm}, X^i$, but also on the moving frame degrees of freedom \\
 ($X^{\underline{m}} = 1/2 X^{+} u^{-\underline{m}} + 1/2 X^{-}
 u^{+\underline{m}} -  X^{i} u^{i\underline{m}}$)  which can be regarded
 as  related to 1-form variables $f^{\pm i}, \omega , A^{ij}$ constrained
 to satisfy (\ref{pMC}).

The e.o.m.-s appearing as a result of variations with respect to the
Cartan forms and the vielbeine $e^{\pm}$ are nondynamical
\begin{equation}\label{eian} E^i \equiv DX^i - 1/2 X^{\pm} f^{\mp i}
= 0 
\end{equation}
\begin{equation}\label{e+an}
E^{+} \equiv
DX^{+} -  X^{i} f^{+i}
= e^{+}
\end{equation}
\begin{equation}\label{e-an}
E^{-} \equiv
DX^{-} -  X^{i} f^{-i}
= e^{-}\; .
\end{equation}
These 'rheotropic' conditions  are similar to the algebraic equations  appearing in the PSM approach \cite{strobl,kummer}.
Finally also (\ref{PC+})--(\ref{G}) must be taken into account. In
addition the forms $f^{\pm i}$  satisfy  (\ref{symm})

$$ f^{-i} \wedge e^{+} + f^{+i}\wedge e^{-}  =  0 $$
or
\begin{equation}
\label{symman} 
 f^{-i}_{-} = f^{+i}_{+} \equiv 1/2\; h^i 
\end{equation}
which  follows from selfconsistency conditions of the rheotropic relations
(\ref{ei}).

The Riemann curvature is determined in terms of $SO(1,D-1)/ (SO(1,1)
\times SO(D-2))$ connection forms $f^{\pm i}$ ($ i= \perp~~for~~
D=3$) by the Gauss equation (\ref{G})

\begin{equation}\label{G2}
{\cal R} = d\omega = 1/2 e^{+} \wedge e^{-} ( f^{~-i}_{+} f^{~+i}_{-} - (h^i)^2 )
\end{equation}
(Here Eq.\ (\ref{symm}) is taken into account, the Ricci tensor which is
proportional to ${}^{*}{\cal R}$ should not be confused with ${\cal R}$ in
(\ref{G2})).

In several respects  the KR field assumes the role of the potential $V$ in the Einstein-Cartan approach for $d=2$ \cite{strobl,kummer}.
Its dependence on moving frame variables (related to the connection forms)
gives the possibility to describe not only matterless gravity, but any two-dimensional gravity in accordance with the general theorems  about local isometric embedding \cite{Ei,Kr} mentioned above.
Still, this approach so far did not achieve an action with fully unconstrained variables and with the expected full PSM structure.

\subsection{Unconstrained Variables and 'Picture Duality'}

In 4.2 we identified the analytical basis coordinates $X^{\pm},
X^{i}$ with the Lagrange multipliers $\tilde{X}$ of the
Einstein-Cartan approach \cite{kummer,strobl}.  Now we show that
the picture could be changed to a 'dual' one containing 
more straightforward counterparts of the $\tilde{X}$ field. 
In fact this 'dual' picture appears when the
Einstein-Cartan approach is incorporated at a more basic level.

We now introduce also the Maurer-Cartan Equations (\ref{PC+})--(\ref{G})
with Lagrange multipliers into the free string action ((\ref{acanf}) with
$B_2=0$) 

\begin{eqnarray}
\label{acanf1} 
S = \int ( - 1/2  X^{+}
{\cal D} e^{-} + 1/2  X^{-} {\cal D} e^{+} +  1/2 e^{+} \wedge e^{-} 
+ \nonumber\\ 
+ X^i (f^{-} \wedge e^{+}- f^{+} \wedge e^{-}) + \nonumber\\ 
+ Y^{+i} {\cal D} f^{-i} + 1/2  Y^{-i} {\cal D} f^{+i} + Y (d\omega - 1/2
f^{-} \wedge f^{+}) \nonumber\\ 
+ Y^{ij} (dA^{ij} + A^{ik} \wedge A^{kj} +
f^{-[i} \wedge f^{+j]})\; .  
\end{eqnarray}

The last two lines of (\ref{acanf1}) have the form of Einstein-Cartan-like
model (\ref{Psm}) without potential, but for the vielbein replaced by the
pull-back of the $SO(1,D-1)/[SO(1,1)\times SO(D-2)]$ coset covariant forms
$f^{\pm i}$. The first and the third line are just the actions for 
two Einstein-Cartan gravity models, and the second line 
and the second line produces an interaction of these two models, gluing them 
together.

In fact, the entire action (\ref{acanf1}) represents a general PSM model
(\ref{PsMg}) with one forms $$ {\cal A}_{\cal B} = \left( e^{\pm}, f^{\pm
i}, \omega, A^{ij} \right).  $$

For the simplest case $D=3$ the number of the forms $f^{\pm}$ is the same as of vielbein forms $e^{\pm}$,
the gauge fields $A$ disappear,  and the structure of the action functional
\begin{eqnarray}\label{acanf1D3}
S =       \int (
- 1/2  X^{+} {\cal D} e^{-}
+ 1/2  X^{-} {\cal D} e^{+}
+  1/2 e^{+} \wedge e^{-}  \nonumber\\
X^{\perp} (f^{-} \wedge e^{+}- f^{+} \wedge e^{-}) \nonumber\\ + Y^{+} {\cal D}
f^{-} + 1/2  Y^{-} {\cal D} f^{+} + Y (d\omega - 1/2  f^{-} \wedge f^{+})
\end{eqnarray}
becomes almost symmetric with respect to replacement $e$ by $f$ and coordinate fields $X$ by  Lagrange multipliers $Y$.

Hence, supposing the forms $f^{\pm}$
to be linear independent, we could consider them in the cotangent space as
an alternative basis to the $e^{\pm}$.
Passing from the $e$-basis to the $f$-basis we thus obtain  another picture of the model, which in some sense is dual to the initial one.
As a consequence of this we will be able to exclude vielbeins $e$ and variables $X^{\pm}$ from the action.

\subsection{Dual picture for free string theory and Jackiw-Teitelboim model}

The e.o.m.-s from variation with respect to the vielbein $e^{\pm}$ are 
purely algebraic 

\begin{equation}\label{al12}
e^- = {\cal D}X^- - X^i f^{-i} , \qquad e^+ = {\cal D}X^+ - X^i f^{+i} , 
\end{equation}
and, hence, can be substituted back into the action so that $e^{\pm}$ are eliminated.
The same is true for the variations with respect to the coordinate fields $X^{\pm}$ 

\begin{equation}\label{al34+}
f^{-i} \wedge ({\cal D} X^i - 1/2 X^- f^{+i} )= 0 
\end{equation}
\begin{equation}\label{al34-}
f^{+i} \wedge ({\cal D} X^i - 1/2 X^+ f^{-i} )= 0 
\end{equation}
Eqs. (\ref{al34+}), (\ref{al34-}) are solved  by
\begin{equation}\label{al34+sol}
X^{-} = { {* (2{\cal D} X^i \wedge f^{-i})}
\over {* (f^{+i} \wedge
f^{-i})} } = {1 \over {\cal R}}  ({\cal D}_+ X^i f_-^{~-i} - {\cal D}_-
X^i f_+^{~-i})
\end{equation}
\begin{equation}\label{al34-sol}
X^{+} = {
{* (2{\cal D} X^i \wedge f^{+i})} \over {* (f^{+i} \wedge f^{-i})}
} = {1 \over {\cal R}}  ({\cal D}_- X^i f_+^{~+i} - {\cal D}_+ X^i
f_-^{~+i})
\end{equation} where $$ {\cal R} \equiv 1/2 (f^{~+i}_{+}
f^{~-i}_{-} - f^{~+i}_{-} f^{~-i}_{+}) $$ can be regarded as the Ricci
curvature scalar divided by the square root of the metric, or as some
counterpart of the determinant of the metric in the dual picture.  E.g. in
the simplest $D=3$ case indeed $$ {\cal R} \equiv 1/2 (f^{~+}_{+}
f^{~-}_{-} - f^{~+}_{-} f^{-i}_{+}) = det \left( \matrix{ f^{~+}_{+} &
f^{-}_{+} \cr f^{~+}_{-} & f^{-}_{-}\cr } \right) $$ follows.
Substituting (\ref{al12}), (\ref{al34+}), (\ref{al34-})  into
(\ref{acanf1}) we arrive at
\begin{equation}\label{acanf2}
S^\prime =
\int ( Y^{ij} (dA^{ij} + A^{ik} \wedge A^{kj} + f^{-[i} \wedge f^{+j]})
\end{equation} 
$$ + 1/2 \tilde{Y}^{+i} {\cal D} f^{-i} +  1/2
\tilde{Y}^{-i} {\cal D} f^{+i} + \tilde{Y} d\omega + 1/2 \tilde{Y} f^{+i}
\wedge f^{-i} $$ $$ +{1\over {2{\cal R}}} ({\cal D}_+ X^j f_-^{~-j} -
{\cal D}_- X^j f_+^{~-j}) ({\cal D}X^i \wedge f^{~+i}) $$ $$ - {1 \over
{2{\cal R}}} ({\cal D}_- X^j f_+^{~+j} - {\cal D}_+ X^j f_-^{~+j})  ({\cal
 D}X^i \wedge f^{~-i}) $$ $$ + {1 \over {4{\cal R}^2}} ({\cal D}_+ X^j
 f_-^{~-j} - {\cal D}_- X^j f_+^{~-j}) \times $$ $$ \times ({\cal D}_-
 X^{j^\prime } f_+^{~+j^\prime } - {\cal D}_+ X^{j^\prime }
 f_-^{~+j^\prime })~~  f^{+i} \wedge f^{-i} $$ $$ + 1/2 X^i X^j f{+i}
 \wedge f^{-j}) $$ where the  redefinitions $$ \tilde{Y} = Y - 1/2 X^+ X^-
 , \qquad $$ 
 \begin{equation}\label{redlm} \tilde{Y}^+ = Y^+ + 1/2 X^+
X^\perp ,  
\end{equation} $$ \tilde{Y}^- = Y^- + 1/2 X^- X^\perp
\qquad $$ of the Lagrange multipliers $Y$ have been performed.

The action (\ref{acanf2}) even with GKR background describes complicated
(not very illuminating) nonminimal gravity interactions of ''pre-matter''
scalar fields $X^i$ and natural gauge fields $A^{ij}$.

The action for the dual situation in the  D=3 case, however, 
becomes simple: 

\begin{equation}\label{acanf2D3}
S^\prime =       \int ( 1/2 \tilde{Y}^{+} {\cal D} f^{-} +  1/2    \tilde{Y}^{-} {\cal D} f^{+}
+ \tilde{Y} d\omega +
\end{equation}

$$
+ 1/2 \tilde{Y} f^{+} f^{-}
- 1/2  f^+ f^- ( \nabla^f_{-} X^\perp \nabla^f_{+} X^\perp + (X^\perp )^2 )) 
$$
Here the $\nabla^f_{\pm}$ are covariant derivatives appearing in the decomposition of the differential on the basis of the cotangent space provided by the forms $f^\pm$
$$
d = d\xi^m \partial_m =
e^{\pm} \nabla_{\pm} = f^{\pm} \nabla^f_{\pm}\; .
$$

The first line of the action (\ref{acanf2D3}) coincides with the
Einstein-Cartan action with constant potential for $d=2$ gravity described
by the vielbein $f^\pm$.  The second line is the minimally coupled action
for a massive 'pre-matter' scalar field interacting with '$f$-gravity'.
Thus this action describes a Jackiw-Teitelboim model \cite{jackt} coupled
to matter which has provided a very useful laboratory for the study of the
quantization for gravity.

 It is important that the dependence of the action functional on the field
 $X^\perp$ is bilinear. So, in a quantum theory, the integration over this
 field still would provide a relatively simple effective action.

 It should be stressed that in the derivation of the action (\ref{acanf2})
 from (\ref{acanf1}) we have used only nondynamical equations. Hence the
 same can be done for the more general case of arbitrary KR background.
 The result will be the sum of the action (\ref{acanf2}) with the same
 interaction term as in (\ref{acstr}). However, to simply  integrate out
 the $X^\perp$ field for an arbitrary KR background will not be possible
in general.

 As a particular example, the complete solution of the e..o.m.-s for the
 action (\ref{acanf2D3}) in a conformal gauge is  presented in Appendix A.

\section{Conclusion and Outlook}

In this paper we have shown that $D$-dimensional $p$-brane
theory interacting with a $(p+1)$-rank antisymmetric tensor
(GKR) field represents a dynamical system providing a model for
description of a general type of $d=(p+1)$ dimensional gravity
in the frame of the isometric embedding formalism
\cite{Ei,emb1,Kr}, if the number of dimensions $D$ of the target
flat space time satisfies $D\geq (p+1)(p+2)/2$.

This has been done using the moving frame (Lorentz harmonic)
formulation \cite{bzst,bzp,bpstv} of the bosonic $p$-brane
theories which produces the master equations of the so-called
geometric approach \cite{lr,barnes,Zh89,zhelt},
\cite{bpstv,rahiv,zero,b1} as e.o.m.-s.  $d$-dimensional
''physical'' matter appears in such models as a manifestation of
the GKR background.

As a simple example we have studied a model for $d=2$ gravity
provided by $D$-dimensional string theory in more detail.  We 
found that the model possesses a PSM structure
\cite{strobl,kummer,kummer1,kummer2} in the general case of arbitrary
$D$.  Also a  deeper relation between a $D=3$ string model with the PSM
action for $2$-dimensional matterless gravity
\cite{strobl,kummer,kummer1} appeared.  The simplest model
of a free bosonic string was shown to be equivalent to a
Jackiw-Teitelboim model \cite{jackt}.

For $d=4$ our model realizes the idea of Regge and Teitelboim
for a 'string-like' description of gravity \cite{regge} and
provides a dynamical ground for description of General
Relativity within the embedding approach \cite{emb1,Kr}.  In our
framework the Universe can be considered as a $3$-brane in
$D=10$ dimensional space-time with a rank-$4$ antisymmetric
tensor background.  Matter in this Universe appears as a
manifestation of $D=10$ GKR field.

It seems to be more than a coincidence that the number of target
space-time dimensions $D=10$ is distinguished as a critical
dimension of superstring theory too, inspiring 
speculations about a relation of the model considered here with
string theory.

As it is well known \cite{hor}, in the type IIB superstring spectrum
a self-dual four-form gauge field appears.  Moreover, among the
so-called string solitons  in IIB superstring theory
there is a 3-brane \cite{hor}.  In accordance with the
Mantonen-Olive conjecture \cite{mantonen}, the dual theory,
where solitons become fundamental objects, should exist.  Such a 
dual theory is just the one of a (Dirichlet $N=2$
super-)3-brane, being under active investigation now
\cite{c0,c1,c2,schw,bt,bst}.  A 4-form GKR gauge field can be coupled
naturally to this 3-brane.  For nontrivial GKR background the
embedding of the 3-brane into flat 10-dimensional Minkowski
space-- time should be nonminimal and should describe arbitrary
curved 4-dimensional Einstein space-time which may be suitable
as a model for the Universe.  Thus a (simplified) model of the
effective action for such a 'solitonic' Universe seems to be
covered by our approach.

It is interesting that the idea of the embedded Universe,
explored here seems to attract renewed interest also from the
field theoretical point of view.  Recently a new study
\cite{shifman} devoted to a dynamical compactification mechanism
appeared.  The idea of a dynamical generation of the Universe as
a 4-dimensional topological (or nontopological) defect in
higher dimensional space-time is its central subject.  The
authors of Ref.\ \cite{shifman} deal with the embedding of low
dimensional models of the Universe in the form of a domain wall
and of a cosmic string and have found that this mechanism can
even help to solve the problem of supersymmetry breaking in the
Universe.

A natural next step could be to consider gravity models inspired
by superstrings and supermembranes.  The existence of
supersymmetric generalizations of the extrinsic geometry
approach \cite{bpstv} and of our model \cite{bsv} make
supersymmetric generalizations of the 'string-like' description
of gravity rather straightforward.

 In high dimensional space-time the only multiplets are the
supergravity ones.  So, there is some hope to overcome the
difficulties from quantum gravity when free super-$p$-branes and
super-$D_p$-branes are taken as point of departure.  Recent
progress in studying $D=10$ Dirichlet super-p-branes
\cite{hs1,c1,c2,schw,bt,bst}, 11-dimensional 5-branes \cite{hs2} and
F-theory \cite{Fth1,Fth2,12D} opens the possibility of new
applications of our approach for ''physical'' supergravity. 
This would correspond to the search for an adequate variant of
the 'preon' model \cite{preons}. Models of that type were very
popular after the realization that nontrivial counterterms for
$N=8$ supergravity may be produced in higher loops (which
destroyed, at least for a time, the hope for finiteness of that
theory) and that the field contents of $N=8$ supergravity is not
sufficient to provide physical gauge fields at low energies (see
\cite{preons} and refs.\ therein).

\section*{Acknowledgements}

One of the authors (I.B.) thanks A.\ Kapustnikov, I.\ Klebanov,
V.\ Nesterenko, A.\ Nurmagambetov, T.\ Ortin, A.\ Pashnev, D.\
Sorokin, K.\ Stelle, Yu.\ Stepanovskij, M.\ Tonin, A.\
Zheltukhin, V.\ Zima for interest in this work and helpful
discussions.

 This work was supported in part by INTAS Grant 93-633-EXT,
INTAS, the Dutch Government Grant No.\ {\bf 94-2317} and the
Austrian Science Foundation (Fonds zur F\"orderung der 
wissenschaftlichen Forschung) project {\bf P-10221-PHY}.

\newpage

\section*{Appendix A: Free string in D = 3.
Complete solution of the equations of motion.}

The free bosonic string in embedding dimension D is the simplest
model for $d=2$ gravity within this approach.  It allows a
complete solution.  The e.o.m.-s following from the action
(\ref{acanf}) split into a set of one-form equations

\begin{equation}\label{1+}
dX^{+} = X^+ \omega + X^\perp f^+ + e^+ \;  , 
\end{equation}
\begin{equation}\label{1-}
dX^{-} = - X^- \omega + X^\perp f^- + e^- \; , 
\end{equation}
\begin{equation}\label{2}
dX^{\perp } = {1 \over 2} X^+ f^- + {1 \over 2} X^- f^+ \; ,
\end{equation}

which naturally produce relations similar to conservation laws
in PSM-models \cite{strobl,kummer} 
\begin{equation}\label{8'}
d(X^{+}X^{-}) = X^+ (e^- + X^\perp f^- ) + X^- (e^+ + X^\perp
f^+ )\; , 
\end{equation}

$$
ds^2 \equiv d(X^{+}X^{-} - (X^\perp)^2) = X^+ e^- + X^- e^+ \; .
$$
The set of  two-form equations comprises \begin{equation}\label{7''}
e^+ \wedge f^- + e^- \wedge f^+ = 0,
\end{equation}
\begin{equation}\label{3}
f^+ \wedge e^- - f^- \wedge e^+ =
e^+ \wedge e^- h(X,u), \qquad
\end{equation}
\begin{equation}\label{6+}
{\cal D} e^+ = d e^+ -
e^+ \wedge \omega = 0, \qquad
\end{equation}
\begin{equation}\label{6-}
{\cal D} e^- = d e^- +
e^- \wedge \omega = 0 \; .
\end{equation}

>From variation of $X^\perp$ follows the 'Gauss' equation 
\begin{equation}\label{4}
d\omega = {1 \over 2} f^- \wedge  f^+
\end{equation}
and from $\delta Y^\pm$ the Peterson-Codazzi equations 
\begin{equation}\label{5+}
{\cal D} f^+ = d f^+ -
f^+ \wedge \omega = 0, \qquad
\end{equation}
\begin{equation}\label{5-}
{\cal D} f^- = d f^- +
f^- \wedge \omega = 0 , \qquad
\end{equation}

\subsection*{A.1 Two-form equations}
The two-form equations  contain the
integrability conditions for the one-form equations and the additional proper dynamical equation
(\ref{3}).  We follow the line of argument presented in \cite{zero,rahiv}.

In this case  (\ref{7''}) and (\ref{3}) result in (cf.\ the 
notation (\ref{fpmi1})) 
$$
f^{~+}_{+} = f^{~-}_{-} = 0,
$$
and hence
\begin{equation}\label{I1+}
f^{~+} = e^{-} f^{~+}_{-} , \qquad
\end{equation}
\begin{equation}\label{I1-}
f^{~-} = e^{+} f^{~-}_{+} , \qquad
\end{equation}
>From (\ref{I1+}), (\ref{I1-}), (\ref{5+}), (\ref{5-}) the expression for
the spin connection form 
\begin{equation}\label{I3} \omega =   {1 \over 2}
e^+ \nabla_+ ln(f^{~+}_{-}) -  {1 \over 2} e^- \nabla_- ln(f^{~-}_{+}) ,
\end{equation} can be obtained.  Substituting (\ref{I3}) into  (\ref{6+}),
(\ref{6-}), one could write the latter as condition for some  invariant
forms to be closed:  
\begin{equation}\label{I4+} d(e^+ (f^{~-}_{+})^{1/2}) = 0 
\end{equation} 
\begin{equation}\label{I4-} d(e^- (f^{~+}_{-})^{1/2}) = 0 
\end{equation} 
For trivial topology of the world sheet  eqs.\
 (\ref{I4+}), (\ref{I4-}) are simply solved by
\begin{equation}\label{I5+} e^+ =  (f^{~-}_{+})^{-1/2})
 g_{(+)}(\xi^{(+)}) d\xi^{(+)} \end{equation} 
 \begin{equation}\label{I5-}
 e^- =  (f^{~+}_{-})^{-1/2}) g_{(-)}(\xi^{(-)}) d\xi^{(-)} 
 \end{equation}
where, in general, $\xi^{(\pm)}=\xi^{(\pm)}(\xi^m)$ are some functions of the world sheet coordinates $\xi^m$ and $g_{(-)}= g_{(-)}(\xi^{(-)})$, $g_{(+)}= g_{(+)}(\xi^{(+)})$ are arbitrary  functions of the $\xi^{(-)}$ and $\xi^{(+)}$ respectively.

It is useful to choose local world sheet coordinates  coincident with the functions $\xi^{(\pm)}$:
$$
\xi^m = (\xi^{(+)}, \xi^{(-)}) .
$$
This (gauge-)choice breaks general coordinate (reparametrization) invariance up to the conformal transformations, whose parameters are expressed by the  arbitrary chiral functions $g_{(+)}$ and $g_{(-)}$ $$
\partial_{(-)} g_{(+)} \equiv
{\partial \over {\partial \xi^{(-)}}} g_{(+)} = 0 , \qquad \partial_{(+)} g_{(-)} \equiv
{\partial \over {\partial \xi^{(+)}}} g_{(-)} = 0 \quad .
$$
Now the expressions (\ref{I1+}), (\ref{I1-}), (\ref{I3}) for the forms
$f^{\pm}, \omega$ become 
\begin{equation}\label{I7+} f^{~+} = e^{-}
f^{~+}_{-} = d\xi^{(-)}~~  g_{(-)}(\xi^{(-)}) ~~(f^{~+}_{-})^{+1/2} ,
\qquad  
\end{equation} 
\begin{equation}\label{I7-} f^{~-} = e^{+}
f^{~-}_{+} =  d\xi^{(+)} ~~g_{(+)}(\xi^{(+)})~~ (f^{~-}_{+})^{-1/2} ,
\qquad 
\end{equation} 
\begin{equation}\label{I8} \omega =    d\xi^{(+)}
\partial_{(+)} ln(f^{~+}_{-})^{1/2} -  d\xi^{(-)} \partial_{(-)}
ln(f^{~-}_{+})^{1/2} \; .  
\end{equation} 
The Gauss equation (\ref{4}) for
the forms (\ref{I8}), (\ref{I7-}), (\ref{I7+}) produces the relation
$$
\partial_{(-)} \partial_{(+)} (f^{~+}_{-} f^{~-}_{+})^{1/2} =
$$
\begin{equation}\label{I9}
 = {1 \over 2} (f^{~+}_{-} f^{~-}_{+})^{1/2} ~g_{(+)}(\xi^{(+)})
 ~~g_{(-)}(\xi^{(-)}) \; .
\end{equation}
Denoting
\begin{equation}\label{I10}
(f^{~+}_{-} f^{~-}_{+})^{1/2} ~~g_{(+)}(\xi^{(+)}) ~~g_{(-)}(\xi^{(-)}) \equiv e^{2W} ,
\end{equation}
eq.\ (\ref{I9}) turns into the nonlinear Liouville equation
\begin{equation}\label{I9'}
\partial_{(-)} \partial_{(+)} W = {1 \over 4}
e^{2W} \end{equation} whose general solution is well-known (see, for
example, \cite{kulish}) 
\begin{equation}\label{I12} 
W = {1 \over 2} ln {
{4 \partial_{(+)} A \partial_{(-)} B} \over {(A+B)^2}}, \qquad
\partial_{(-)} A = 0 = \partial_{(+)} B \; .
\end{equation}
Since this equation is typical for the 2d gravity with constant
curvature its appearance is not surprising, because the related
Jackiw-Teitelboim model (without matter) precisely has this property.

\subsection*{A.2: Solution of One-Form Equations}

Eq. (\ref{2}) may be rewritten as
$$
dX^\perp = 1/2 X^+ f^- + 1/2 X^- f^+ =
$$
\begin{equation}\label{I13}
 1/2 d\xi^{(+)} g_{(+)} (f^{~-}_{+})^{1/2} X^+
 \end{equation}
$$
+ 1/2 d\xi^{(-)} g_{(-)}
 (f^{~+}_{-})^{1/2} X^-
$$
which means that the  $X^{\pm}$ fields are expressed through the derivatives of $X^\perp $ field
\begin{equation}\label{I14+}
X^+ = 2 (f^{~+}_{-})^{-1/2} (g_{(+)})^{-1} \partial_{(+)}X^\perp
\end{equation}
\begin{equation}\label{I14-}
X^- = 2 (f^{~-}_{+})^{-1/2} (g_{(-)})^{-1} \partial_{(-)}X^\perp \; .
\end{equation}
Since Eq.\ (\ref{1+}) expressed  in terms of coordinates $\xi^{\pm}$
becomes \begin{equation}\label{I15+'} (f^{~+}_{-})^{1/2} \partial_{(+)} (
(f^{~+}_{-})^{-1/2} X^+) = g_{(+)} (f^{~-}_{+})^{-1/2} \end{equation}
\begin{equation}\label{I16+'}
(f^{~-}_{+})^{-1/2} \partial_{(-)}
( (f^{~-}_{+})^{+1/2} X^+) = g_{(-)} (f^{~+}_{-})^{1/2} X^\perp \; ,
\end{equation}
substituting in (\ref{I14+}) yields
\begin{equation}\label{I15}
\partial_{(+)}(e^{-2W} \partial_{(+)} X^\perp ) = 1/2 (g_{(+)})^2 e^{-2W} 
\end{equation}
\begin{equation}\label{I16}
\partial_{(-)}\partial_{(+)} X^\perp = 1/2 X^\perp e^{-2W} \end{equation}
where the definition (\ref{I5-}) of  $W$ has been  used.

Eq.\ (\ref{I12})  for $W$ in eq.\ (\ref{I16}) yields 

\begin{equation}\label{I16'}
\partial_{(-)} \partial_{(+)} X^\perp  =  2 X^\perp {{ \partial_{(+)} A \partial_{(-)} B}
\over {(A+B)^2}} ,
\end{equation}
and the general solution of the {\sl linear} equation (\ref{I16'}) is
simply 
\begin{equation}\label{I17} X^\perp = C_1 {1 \over {A+B}} + C_2
(A+B)^2, \qquad \partial_{(-)} A = 0 = \partial_{(+)} B 
\end{equation}
where $C_1, C_2$ are two integration constants. Substituting the solution
(\ref{I17}) and the expression (\ref{I12}) for the superfield $W$ into
(\ref{I15+'}) we reduce the latter to the expression for the chiral
function $g_{(+)}= g_{(+)}(\xi^{(+)})$ in terms of the chiral function $A=
A(\xi^{(+)})$  which was  the parameter of the general solution of the
nonlinear Liouville equation 
\begin{equation}\label{I18+}
 g_{(+)} = \pm (3C_2)^{1/2} \partial_{(+)} A \; .
\end{equation}
The same procedure can be carried through for the equations involving 
the $X^-$ coordinates.

As a result all the coordinates become expressed in terms of the
chiral functional parameters $A(\xi^{(-)}), B(\xi^{(+)})$ of the
general solution (\ref{I12}) of the Liouville equation, two
integration constants $C_{1,2}$ and the functional parameter
$L=L(\xi^{(\pm )})$ of the gauge $SO(1,1)$ (world sheet Lorentz)
transformations.

So the complete solution of the free string model involves two
constants ('Casimir-functions') of the PSM structure
\cite{strobl}

$$
C_1 =const \qquad  C_2 =const
$$
and two chiral (left-moving and right-moving) functions
$A(\xi^{(+)}), B(\xi^{(-)})$

$$
\partial_{(-)} A = 0 = \partial_{(+)} B $$
involved in the general solution (\ref{I12}) of the Liouville equation.

The complete solution is represented by the following set of relations:
\begin{itemize}
\item
Coordinates of the analytical basis:
$$
X^+ = e^{-L} (\partial_{(+)} A / \partial_{(-)} B )^{1/2}
(- {{C_1}  \over {A+B}} + 2 C_2 (A+B)^2 ), $$
$$
X^- = e^{L} (\partial_{(-)} B /\partial_{(+)} A )^{1/2}
(- {{C_1}  \over {A+B}} + 2 C_2 (A+B)^2 ), $$
$$
X^\perp = {{C_1} \over {A+B}} + C_2 (A+B)^2, $$
\item
Vielbeine:

$$
e^+ = 6 e^{-L} C_2 (A+B)
(\partial_{(+)} A / \partial_{(-)} B )^{1/2} d\xi^{(+)}
$$
$$
e^- = 6 e^{L} C_2 (A+B) (\partial_{(-)} B / \partial_{(+)} A
)^{1/2} d\xi^{(-)}
$$
\item
Pull-backs of the vielbeine of the coset: $SO(1,2)/SO(1,1)$
$$
f^+ = 2 e^{-L} { {(\partial_{(+)} A / \partial_{(-)} B )^{1/2}}
\over {(A+B)}} d\xi^{(-)}
$$
$$
f^- = 2 e^{L}
{ {(\partial_{(+)} A / \partial_{(-)} B )^{1/2}} \over {(A+B)}}
d\xi^{(+)}
$$
\item
Riemann curvature two-form:

$$
R=d \omega = 1/2 f^- \wedge f^+ =
$$
$$
= 2 d\xi^{(+)} \wedge d\xi^{(-)}
{ {(\partial_{(+)} A / \partial_{(-)} B )} \over {(A+B)}}
$$
\item
Metric:

$$
ds^2 = e^+ \otimes e^- = 36 (C_2)^2 d\xi^{(+)} \otimes d\xi^{(-)} (A+B)^2 $$

\end{itemize}

\end{document}